\documentclass{aa}  
\usepackage{natbib}
\bibpunct{(}{)}{;}{a}{}{,} 
\usepackage{graphicx}
\usepackage{float}
\usepackage{xcolor}
\usepackage{graphicx}
\usepackage{dcolumn}
\usepackage{bm}
\usepackage{romannum}
\usepackage{amsmath}
\usepackage{graphicx}
\usepackage{xcolor}
\usepackage[caption=false]{subfig}
\usepackage{booktabs}   
\usepackage{relsize}
\usepackage{sidecap}
\usepackage{graphicx}
\usepackage{stfloats}
\usepackage{placeins}

\usepackage{txfonts}

\begin{document}

\title{Strong Stellar Diffusion from Wave DM Cosmological Simulation and Potential Unified Origin for dSphs, UFGs, and UDGs }





   \author{A. Pozo
          \inst{1}
          \and
          T. Broadhurst\inst{1,2,7}
          \and
          J. Zhang\inst{1}
          \and
          M. Oguri\inst{3}\textdagger
          \and
          K. Umetsu\inst{3}
          \and
          R. Emami\inst{5}
          \and
          L. Hernquist\inst{5}
          \and
          P. Mocz\inst{6}
          \and
          M. Vogelsberger\inst{12}     
          }

  \institute{DIPC, Basque Country UPV/EHU, E-48080 San Sebastian, Spain\\
              \email{alvaro.pozolarrocha@bizkaia.eu; tom.j.broadhurst@gmail.com;}
         \and
             University of the Basque Country UPV/EHU,Department of Theoretical Physics, Bilbao, E-48080, Spain
         \and
             ASIAA, Taipei 10617, Taiwan
         \and
             Department of Physics, National Taiwan University, Taipei 10617, Taiwan
         \and
             Center for Astrophysics $\vert$ Harvard $\&$ Smithsonian, 60 Garden Street, Cambridge, MA 02138, USA
          \and
             Center for Computational Astrophysics, Flatiron Institute, 162 5th Ave, New York, NY 10010, USA
         \and
             Ikerbasque, Basque Foundation for Science, Bilbao, E-48011, Spain
         \and
             Hong Kong University of Science and Technology, Institute for Advanced Study and Department of Physics, IAS TT $\&$ WF Chao Foundation Professor,  Hong Kong
         \and
             Energetic Cosmos Laboratory, Nazarbayev University, Nursultan, Kazakhstan
         \and
             Paris Centre for Cosmological Physics, APC, AstroParticule et Cosmologie, Universit\'{e} de Paris, CNRS/IN2P3, CEA/lrfu,Universit\'{e} Sorbonne Paris Cit\'{e}, 10, rue Alice Domon et Leonie Duquet,  75205 Paris CEDEX 13, France  Emeritus
        \and
             Physics Department $\&$ LBNL, University of California at  Berkeley CA 94720 {\it Emeritus}
        \and
             Department of Physics, Kavli Institute for Astrophysics and Space Research, Massachusetts Institute of Technology, Cambridge, MA 02139, USA
             }

\abstract{
Our $\psi$DM simulations predict stars diffuse throughout dark matter halos over the Hubble time as a random walk, scattered by wave perturbations inherent to $\psi$DM. These diffusing stars locally follow a Gaussian profile (Sersic index $n=0.5$), by the central limit theorem, expanding as $R_{1/2}(t) \simeq ({\hbar/{m_\psi}})^{0.5} \sqrt{t}$ matching well typical $\psi$DM dsph galaxies core-halo profiles. Furthermore, our simulations show that depending on the mass of the halo and the resulting soliton, the strength of the scattering effect can vary significantly, naturally producing progressively more diffuse stellar systems in more massive halos. The observed continuity between faint dwarfs, compact dwarf spheroidal galaxies and ultra-diffuse galaxies can therefore be interpreted as age related, with less diffusion in later-forming dwarfs so they appear smaller and with higher surface brightness today. This stellar scattering would be produced by the random walk of the soliton, which gradually pulls stars from the dense inner core into the outer halo, creating the extended stellar envelopes observed around Local Group dwarfs and increasingly diffuse systems at larger masses. Rather than being a phenomenon restricted only to ghostly looking Ultra Diffuse'' galaxies (UDGs), stellar scattering induced by Wave Dark Matter may provide a unified physical mechanism capable of explaining the structure of most observed galaxies across an enormous mass range, from the faintest ultra-faint dwarfs with stellar masses of $\sim10^7\,M_\odot$ to the most massive and extended UDGs approaching $\sim10^{11}\,M_\odot$. In this picture, the diversity of galaxy sizes and surface brightness profiles emerges naturally from the same underlying wave-driven diffusion process, without requiring distinct formation channels or invoking failed galaxy'' scenarios. The widespread diffuse stellar halos and globular cluster distributions uncovered by recent surveys such as Euclid may therefore represent direct observational signatures of the intrinsic granular dynamics of wave dark matter. }


   \maketitle

\section{Introduction}

The nature of Dark Matter (DM) remains one of the main open questions in cosmology. While the standard Cold Dark Matter (CDM) paradigm successfully explains large-scale structure formation, tensions at galactic and sub-galactic scales have motivated the exploration of alternative models \citep{Moore:1994,Klypin:1999,deBlok:2010,Marsh:2014,Safarzadeh:2021}. Among these, Wave Dark Matter (also known as Fuzzy Dark Matter, FDM) describes DM as an ultra-light bosonic field, typically associated with axion-like particles of mass $m_\psi \sim 10^{-22},$eV, behaving as a coherent Bose-Einstein condensate on galactic scales \citep{Schive:2014,Schive:20142,Schive:2016}.

Due to its extremely low particle mass, Wave Dark Matter exhibits macroscopic quantum effects with kiloparsec-scale de Broglie wavelengths, naturally generating interference patterns and solitonic cores within dark matter halos. Cosmological simulations predict that every halo hosts a central soliton surrounded by fluctuating wave interference structures \citep{Schive:20142}. These solitons possess flat-core density profiles and follow scaling relations with halo mass, becoming denser and more compact in more massive systems.

At scales larger than the de Broglie wavelength, Wave Dark Matter reproduces the successful predictions of CDM, remaining consistent with observations of the Cosmic Microwave Background and large-scale structure formation. However, on galactic scales, the intrinsic wave dynamics introduce new gravitational phenomena absent in standard CDM, potentially affecting stellar kinematics, dwarf galaxy evolution, and the formation of diffuse stellar structures.

In recent years it has become increasingly clear that low-mass dwarf galaxies in the Local Group possess extended stellar and dark matter halos reaching several kiloparsecs \cite{Chiti:2021,Collins:2021,Sato:2025}, challenging the classical picture of dwarfs as compact systems. Such extended structures have now been identified in a wide variety of Local Group galaxies, including both dwarf spheroidals and ultra-diffuse systems, as shown in recent studies by \cite{Pozo:2020,Pozo:2023,Pozo:20242}. Remarkably, many of these galaxies appear to share a common core--halo structure that resembles the density profile predicted for Wave Dark Matter halos, while simultaneously exhibiting a smooth transition between the inner stellar core and the outer diffuse stellar halo.

In this work, we present simulations showing how such structures can naturally arise from the stochastic motion of the soliton embedded within every Wave Dark Matter halo. We exploit the random-walk behaviour associated with the central soliton that generically emerges in Wave Dark Matter ($\psi$DM) halos to explain several universal properties observed in low-mass galaxies. In this framework, time-dependent oscillations and displacements of the soliton induce repeated gravitational scattering events that progressively transfer stellar orbits from the dense central regions toward larger radii, a process commonly referred to as soliton-driven random walk \citep{Schive:2020,Li:2021}.

We present several novel results. First, this mechanism naturally redistributes stars from the inner regions toward the outer halo, producing stellar cores that are ubiquitously observed in dwarf galaxies and are in tension with the cuspy density profiles predicted by standard cold dark matter (CDM) models \citep{deBlok:2010}. Second, the same process modifies the stellar density profile by softening the central density while simultaneously enhancing the stellar density at larger radii, leading to the emergence of a characteristic core–halo structure \citep{Pozo:2024,Pozo:20242,Pozo:2025}. This behaviour closely resembles the universal structural properties observed in Local Group dwarf galaxies.

Finally, for more massive haloes, with total masses comparable to those inferred for ultra-diffuse galaxies ($M_{\rm halo} \sim 10^{11},M_\odot$), the soliton mass and amplitude are sufficiently large that the random-walk mechanism becomes significantly more efficient. In this regime, enhanced stellar scattering can give rise to extended, low-surface-brightness cores and diffuse stellar profiles, providing a natural explanation for the structural properties of UDGs within the $\psi$DM framework. The resume of the whole thesis can be found in figure \ref{Fig:thesis}.

\begin{figure*}[!b]
    \centering
 \includegraphics[width=0.8\textwidth,height=16cm]{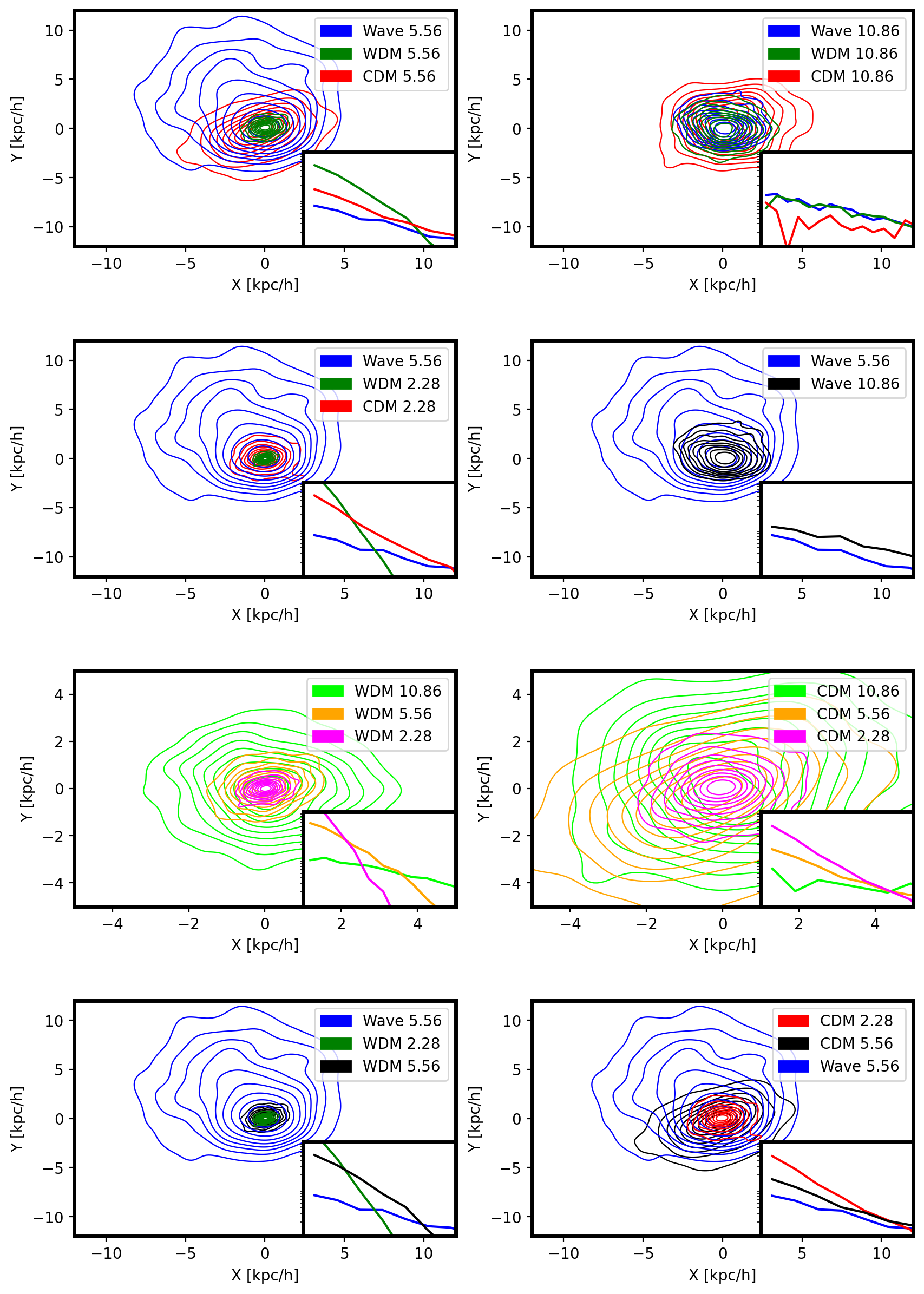}
 \includegraphics[width=1\textwidth,height=5cm]{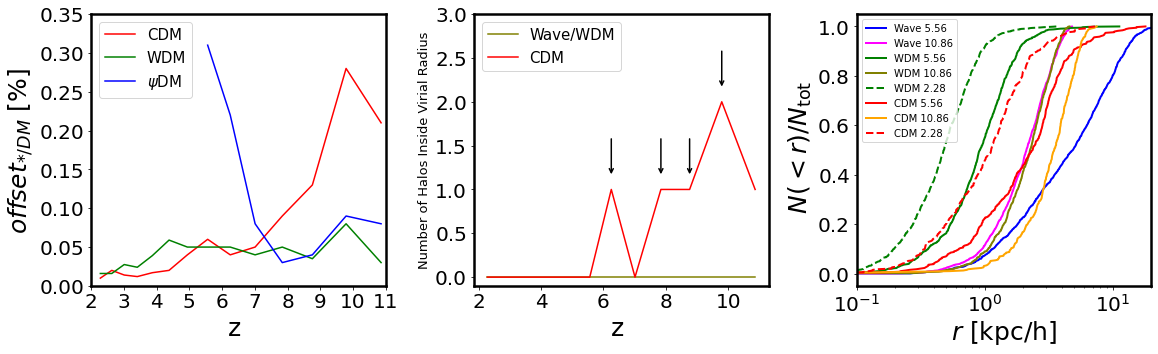}

   \caption{Evolution of the stellar distribution in the most massive simulated galaxy for $\psi$DM, WDM'', and CDM in comoving coordinates. The first four rows show the stellar density evolution and corresponding radial profiles. While CDM and ``WDM'' become increasingly concentrated with time, the $\psi$DM galaxy develops a progressively more diffuse and extended stellar distribution due to soliton-driven scattering. The fifth row shows the stellar--halo offset, merger history, and the cumulative stellar radial distribution, $N(<r)/N_{\rm tot}$, highlighting the outward diffusion of stars driven by the stochastic motion of the central soliton in $\psi$DM.}
    \label{Fig:scatterCOMO}
\end{figure*}

\begin{figure*}
    \centering
 \includegraphics[width=0.8\textwidth,height=16cm]{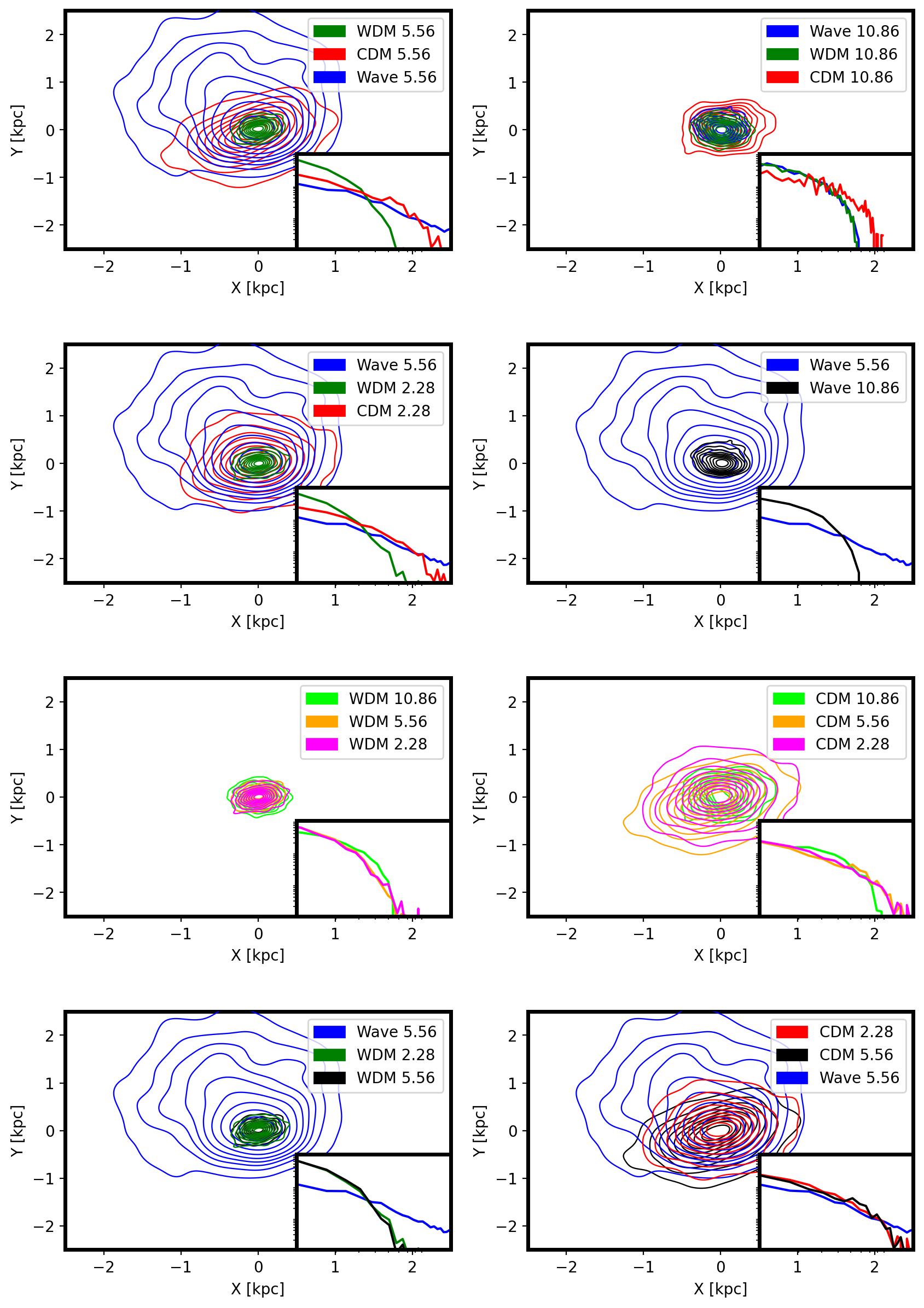}
 \includegraphics[width=1\textwidth,height=5cm]{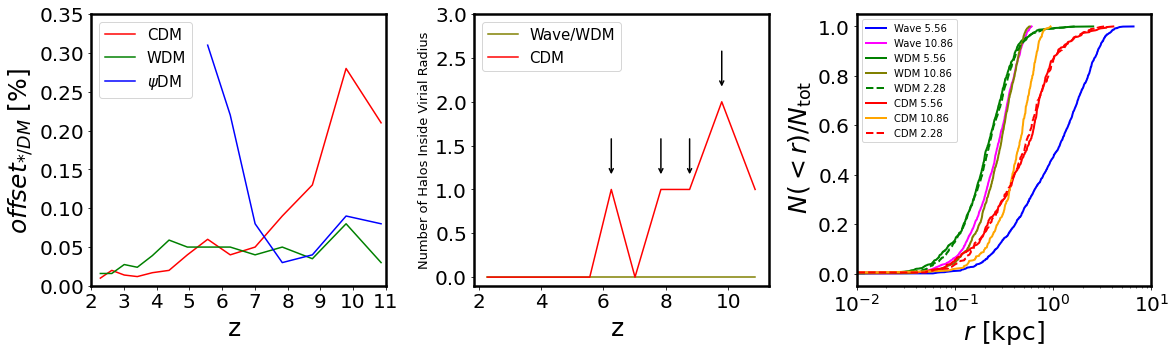}

   \caption{Same as Figure~\ref{Fig:scatterCOMO}, but shown in physical units.}
    \label{Fig:scatterPHY}
\end{figure*}

\section{The wave/fuzzy dark matter halo}
\label{“WDM”-model}
Ultralight bosons, such as ALPs (axion-like particles), have been considered as an ingredient in describing the $\psi$DM \citep{Widrow:1993,Hu:2000}. In the absence of any self-interactions, the boson mass is the only free parameter that describes the DM. For a sufficiently light boson mass, the de-Broglie wavelength exceeds the mean free path, which  leads the ALP to satisfy the ground state condition for a Bose-Einstein condensation. The governing equation for $\psi$DM is the
Schr\"{o}dinger-Poisson (SP) equation, which in comoving coordinates reads as:
\begin{align}
& \biggl[i\frac{\partial}{\partial \tau} + \frac{\nabla^2}{2} - aV\biggr]\psi=0\,,\\
& \nabla^2 V =4\pi(|\psi|^2-1)\,.
\end{align}
where $\psi$ is the wave function (see \cite{Schive:2014} for the explanation of the interpretation of the wave function, originally introduced by \cite{Widrow:1993}, to describe the mean field behavior of the wave function in this context), $V$ is the gravitation potential and $a$=(1/(1+z)) is the cosmological scale factor, where z is the redshift.The comoving length $\bm{x}$ is normalized to ($\frac{3}{8\pi} H_0^2 \Omega_{m0})^{-1/4}$ $(m_\psi/\hbar)^{1/2}$, the time normalized to $d\tau=(\frac{3}{8\pi} H_0^2 \Omega_{m0})^{1/2} a^{-2}dt$ and the wave function $\psi$ normalized to $(\rho_{m0}/m_{\psi})^{1/2}$, where $H_0$ is the present Hubble parameter,$\Omega_{m0}$ is the matter density parameter and $\rho_{m0}$ the background mass density, \citep{Schive:20142}. Finally, the physical interpretation of the wave function in $\psi$DM is that the density is given by $\rho = |\psi|^2$.

Recently, it has become possible to perform high dynamic range cosmological simulations that solve the above equations \citep{Schive:2014, Schwabe:2016, Mocz:2017, May:2021}, where GPU computing has allowed an efficient adaptive mesh refinement \citep{Schive:2014}. These evolve to produce large-scale structures indistinguishable from CDM, but with virialized halos characterized by a solitonic core in the ground state that naturally explains the dark matter-dominated cores of dSph galaxies, previously analyzed in many dSph and UFD of the Local Group \citep{Pozo:2020,Pozo:2023,Pozo:2021,Pozo:2022,Schive:20142} . Another important feature arising from the simulations is that the central soliton is surrounded by an extended halo with a granular texture on the de-Broglie scale, due to interference of excited states, but which when azimuthally averaged follows the Navarro-Frenk-White (NFW) density profile \citep{Navarro:1996,Woo:2009,Schive:2014,Schive:20142}.

The fitting formula for the density profile of the solitonic core in a $\psi$DM halo  is obtained from cosmological simulations \citep{Schive:2014,Schive:20142} as
\begin{equation}\label{eq:sol_density}
\rho_c(r) \sim \frac{1.9~a^{-1}(m_\psi/10^{-23}~{\rm eV})^{-2}(r_c/{\rm kpc})^{-4}}{[1+9.1\times10^{-2}(r/r_c)^2]^8}~M_\odot {\rm pc}^{-3},
\end{equation}
where $m_\psi$ is the boson mass
and $r_c$ is the solitonic core radius. The latter scales with both the halo mass and the boson mass, obeying the following scaling relation which was derived from the simulations (Eq. 7 \citep{Schive:20142})
\begin{equation}\label{eq:sol_radius}
r_c=1.6\biggl(\frac{10^{-22}}{m_\psi}  eV \biggr)a^{1/2}
\biggl(\frac{\zeta(z)}{\zeta(0)}\biggr)^{-1/6}
\biggl(\frac{M_h}{10^9M_\odot}\biggr)^{-1/3} {\rm kpc},
\end{equation}
where $M_h$ is the halo mass and $\zeta(z) \equiv \frac{18 \pi^2 + 82 (\Omega_m(z) - 1) - 39 (\Omega_m(z) - 1)^2}{\Omega_m(z)} \sim 350 \, (180) \text{ at } z = 0 \, (z \geq 1)$. $\Omega_m(z)$ is defined as the matter density parameter as a function of redshift, normalized to the critical density at that time. Beyond the soliton, at radii larger than a transition radius ($r_t$), the simulations also reveal that the halo roughly resembles NFW in shape, presumably reflecting the nonrelativistic nature of condensates beyond the de Broglie scale, and therefore the total density profile can be written as
\begin{equation}\label{eq:dm_density}
\rho_{DM}(r) =
\begin{cases} 
\rho_c(r)  & \text{if \quad}  r< r_t, \\
\frac{\rho_0}{\frac{r}{r_s}\bigl(1+\frac{r}{r_s}\bigr)^2} & \text{otherwise},
\end{cases}
\end{equation}

where $\rho_0$ is chosen such that the inner solitonic profile matches the outer NFW-like profile at approximately $\simeq r_t$, and $r_s$ is the scale radius. In detail, the scale radius of the solitonic solution, which represents the ground state of the Schr\"{o}dinger-Poisson equation, is related to the size of the halo through the uncertainty principle.

\section{Random-walk soliton dynamics in wave dark matter}

Unlike in collisionless cold dark matter, the central solitonic core predicted in $\psi$DM haloes does not constitute an exact stationary equilibrium configuration in realistic cosmological environments. High-resolution simulations show that the core exhibits persistent oscillatory behaviour and stochastic displacements. \citet{Veltmaat:2018} demonstrated that the central density peak undergoes long-lived quasi-periodic oscillations, while \citet{Schive:2020} established that its centre performs a random walk within the host halo potential. This dynamical behaviour is a direct consequence of interference effects inherent to the Schrödinger--Poisson (SP) system.

In the non-relativistic regime, $\psi$DM is governed by
\begin{equation}
i\hbar \frac{\partial \psi}{\partial t}
= -\frac{\hbar^2}{2m} \nabla^2 \psi + m \Phi \psi,
\end{equation}
\begin{equation}
\nabla^2 \Phi = 4\pi G m |\psi|^2,
\end{equation}
where $\psi$ is the wavefunction, $m$ the particle mass, and $\Phi$ the gravitational potential. On timescales short compared to halo evolution, the gravitational potential can be approximated as quasi-static, allowing the wavefunction to be decomposed into a superposition of energy eigenstates \citep{Lin:2018}. The ground state corresponds to the solitonic configuration and dominates the density in the central region. However, excited states also contribute non-negligibly to the total density field. Interference between the ground state and higher-energy eigenmodes naturally reproduces the oscillatory behaviour and stochastic wandering of the density maximum observed in simulations \citep{Li:2021, Padmanabhan:2021}. The soliton is therefore more accurately described as a dynamically evolving interference pattern rather than as an exact stationary solution embedded in a cosmological halo.

In the Madelung \citep{Madelung:1927} representation, writing
\begin{equation}
\psi = \sqrt{\rho/m}\, e^{iS/\hbar},
\end{equation}
the SP system can be recast into fluid-like equations with an additional quantum pressure term,
\begin{equation}
\mathbf{a}_Q = -\nabla Q,
\end{equation}
\begin{equation}
Q = -\frac{\hbar^2}{2m^2}
\frac{\nabla^2 \sqrt{\rho}}{\sqrt{\rho}}.
\end{equation}
Interference between eigenmodes induces spatial and temporal variations in the phase $S$, leading to fluctuations in $\nabla S$ and consequently in the quantum pressure contribution to the acceleration field \citep{Hui:2017,Mocz:2017}. These fluctuations act as a stochastic forcing term on the soliton, producing a diffusive displacement of its centre within the inner halo.

The amplitude and characteristic timescale of the soliton random walk depend on both halo mass and particle mass. In more massive haloes, the enhanced granularity of the surrounding wave field and the increased core mass fraction lead to stronger stochastic forcing, resulting in larger root-mean-square displacements of the soliton centre. Numerical studies report correlations between the magnitude of this displacement and global halo properties such as virial mass and concentration \citep{Li:2021,Schive:2020}.

This dynamical behaviour has important implications for baryonic tracers embedded in the inner halo\citep{Schive:2020}. The time-dependent gravitational potential generated by the wandering soliton induces cumulative perturbations to stellar and gaseous orbits. Over multiple dynamical times, this process leads to orbital diffusion and radial redistribution of baryons, potentially flattening the central stellar density profile and contributing to the formation of extended, low surface-brightness systems. Within the $\psi$DM framework, soliton random-walk dynamics thus provide a physically motivated mechanism for the emergence of cored stellar distributions without invoking strong baryonic feedback or environmental interactions.

\begin{figure*}
    \centering
 \includegraphics[width=1\textwidth,height=4cm]{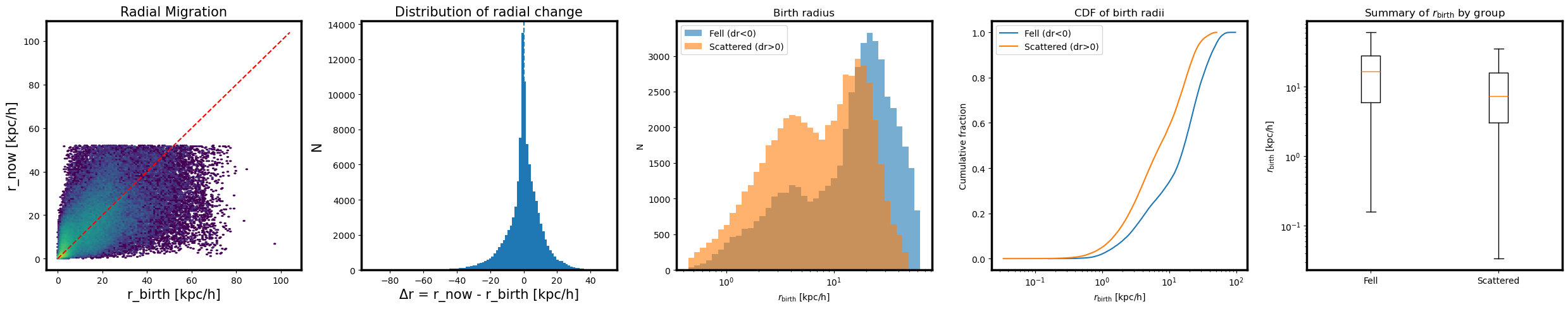}
 \includegraphics[width=1\textwidth,height=4cm]{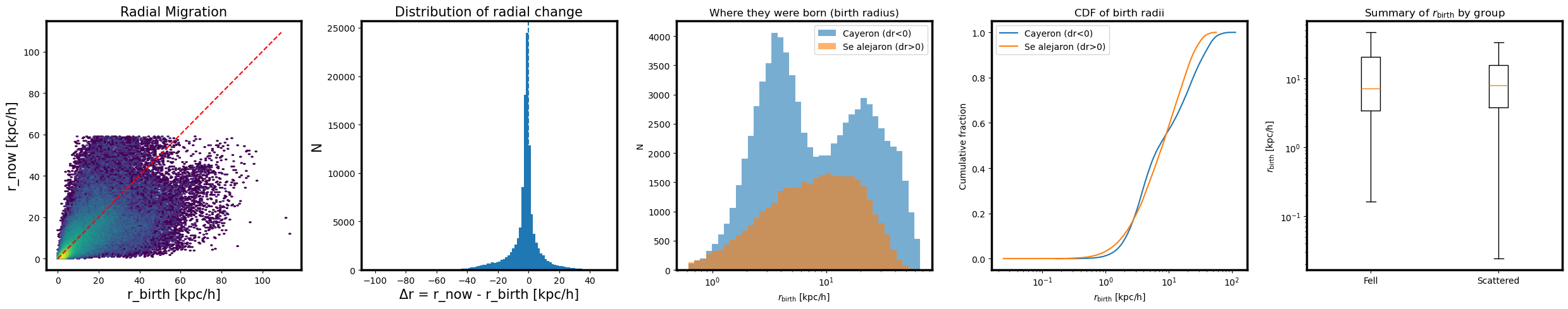}
 \includegraphics[width=1\textwidth,height=4cm]{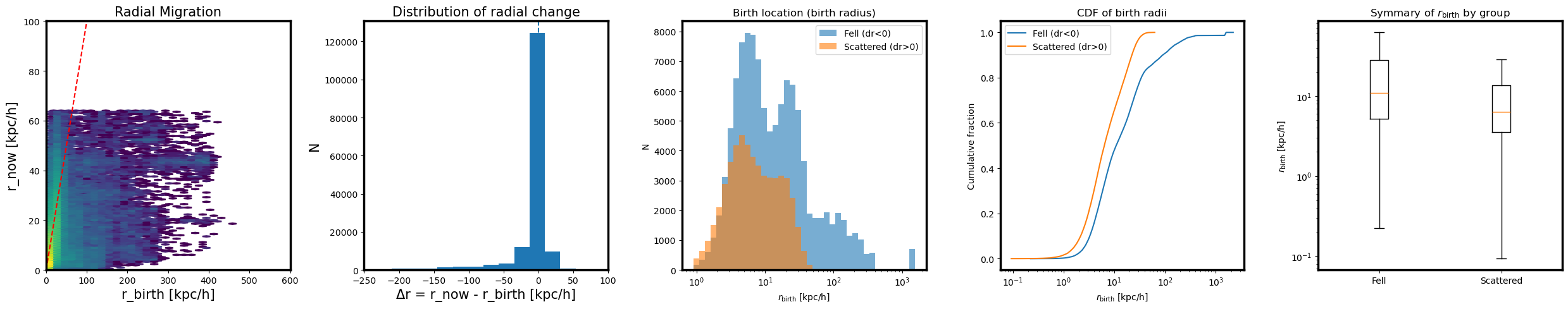}

   \caption{Radial migration properties of stars in the most massive simulated galaxy at $z=5.56$ for $\psi$DM (top row), ``WDM'' (middle row), and CDM (bottom row), shown in comoving units. The panels show: present-day radius versus birth radius, the distribution of radial changes $\Delta r = r_{\rm now}-r_{\rm birth}$, the birth-radius distributions of inward and outward migrating stars, their cumulative distribution functions, and the statistical summary of the birth radii. Only the $\psi$DM simulation exhibits strong outward stellar diffusion, with stars born in the central regions being efficiently scattered toward larger radii due to stochastic perturbations generated by the soliton random walk. In contrast, CDM and ``WDM'' remain largely concentrated around the $r_{\rm now}=r_{\rm birth}$ relation, showing only limited outward migration and no significant stellar diffusion.}
    \label{Fig:RadialMigration}
\end{figure*}

\section{Stellar core evolution -scattering vs contraction}

We base our analysis on the cosmological simulations originally presented by \cite{Mocz:2019}. These simulations follow the evolution of the same cosmological volume under three different dark matter models: CDM, ``WDM'' and $\psi$DM. The so-called ``WDM'' run is not a pure warm dark matter simulation, but rather a CDM simulation initialized with a Wave Dark Matter cutoff in the initial power spectrum, intended to serve as an approximate proxy for $\psi$DM without including its full wave dynamics. The simulations evolve a periodic cosmological box of size $L_{\rm box} = 1.7 \, h^{-1} \, {\rm Mpc}$, assuming a boson mass of $m_\psi = 2.5 \times 10^{-22} \, {\rm eV}$. At this mass scale, the uncertainty principle introduces a suppression in the primordial power spectrum below $L_{\rm cutoff} \approx 1.4 \, h^{-1} \, {\rm Mpc}$. The evolution starts at redshift $z = 127$, corresponding to a Universe age of approximately $10^7 \, {\rm years}$, and continues until $z = 5.5$, when the Universe reaches an age of roughly $10^9 \, {\rm years}$. The final redshift is limited by numerical resolution requirements to ensure fully converged results.

The $\psi$DM simulation employs a spectral resolution of $1024^3$ for the dark matter field and $512^3$ baryonic particles, corresponding to a baryonic mass resolution of approximately $2.64 \times 10^3 \, M_\odot$. Cosmological parameters are adopted from the \textit{Planck} measurements \citep{Planck:2016}, although the normalization amplitude is enhanced from $\sigma_8 = 0.8$ to $\sigma_8 = 1.4$, following \cite{Naoz:2012}, in order to compensate for the relatively small cosmological volume and boost the formation of resolved structures. For comparison, both the CDM and ``WDM'' simulations are evolved with $512^3$ dark matter particles. The ``WDM'' particle mass is chosen to reproduce the same cutoff scale as the $\psi$DM model, corresponding to an effective thermal relic mass of $m_{\rm WDM} \sim 1.4 \, {\rm keV}$. This simulation suite produces halos spanning virial masses in the range $5\times10^7 - 5\times10^{10}\,M_{\odot}$, with stellar masses between $5\times10^6$ and $5\times10^8\,M_\odot$, somewhat lower than those inferred from the observational compilation of \cite{Pandya:2024}.

In this section we present the evolution of the stellar core--halo structure in the most massive galaxy of the simulation. Figures \ref{Fig:scatterCOMO} and \ref{Fig:scatterPHY} show, in the first four rows, the evolution of the stellar distribution over the full available redshift range, from $z=10.86$ to $z=5.56$ for $\psi$DM, and from $z=10.86$ to $z=2.28$ for CDM and ``WDM'', in comoving and physical units respectively. These kernel-density contour maps highlight the fundamentally different evolution of the stellar component in the three dark matter models. In both CDM and ``WDM'', the stellar core progressively becomes more concentrated with time, as gravity dominates the stellar dynamics and drives stars toward the central regions of the halo. In contrast, the $\psi$DM galaxy exhibits the opposite behaviour: the stellar distribution at $z=5.56$ is significantly more extended than at $z=10.86$, indicating a continuous expansion and diffusion of the stellar core over time. To facilitate comparison of the stellar profile evolution, small subpanels showing the radial surface density profiles are included within each of the contour panels.

The fifth row presents three additional panels that provide further context for the previous evolution maps. The left panel shows the offset between the dark matter halo center and the stellar center. In the ``WDM'' simulation, where neither mergers nor a central soliton are present, this offset \citep{Pozo:2023} remains negligible and approaches zero at late times. A similar behaviour is observed in CDM, except at high redshift, where the frequent mergers \citep{Mocz:2020,Pozo:2023} visible in the central panel temporarily displace the stellar component relative to the halo center. Once the merger activity decreases, the stellar and dark matter centers rapidly converge. In contrast, the $\psi$DM halo displays a rapidly increasing offset below $z\sim8$, despite the galaxy remaining fully isolated and free from significant mergers or tidal interactions. This displacement is produced by the stochastic motion of the soliton itself, whose random walk continuously perturbs and scatters the stellar core \citep{Schive:2020,Li:2021}. Finally, the right panel of the last row quantifies the same effect visible in the contour maps: $\psi$DM is the only model in which the stellar component evolves from a compact central distribution into a diffuse and extended structure, in clear contrast with the increasingly concentrated stellar profiles found in CDM and ``WDM''.

\begin{figure*}
    \centering
 \includegraphics[width=1\textwidth,height=9cm]{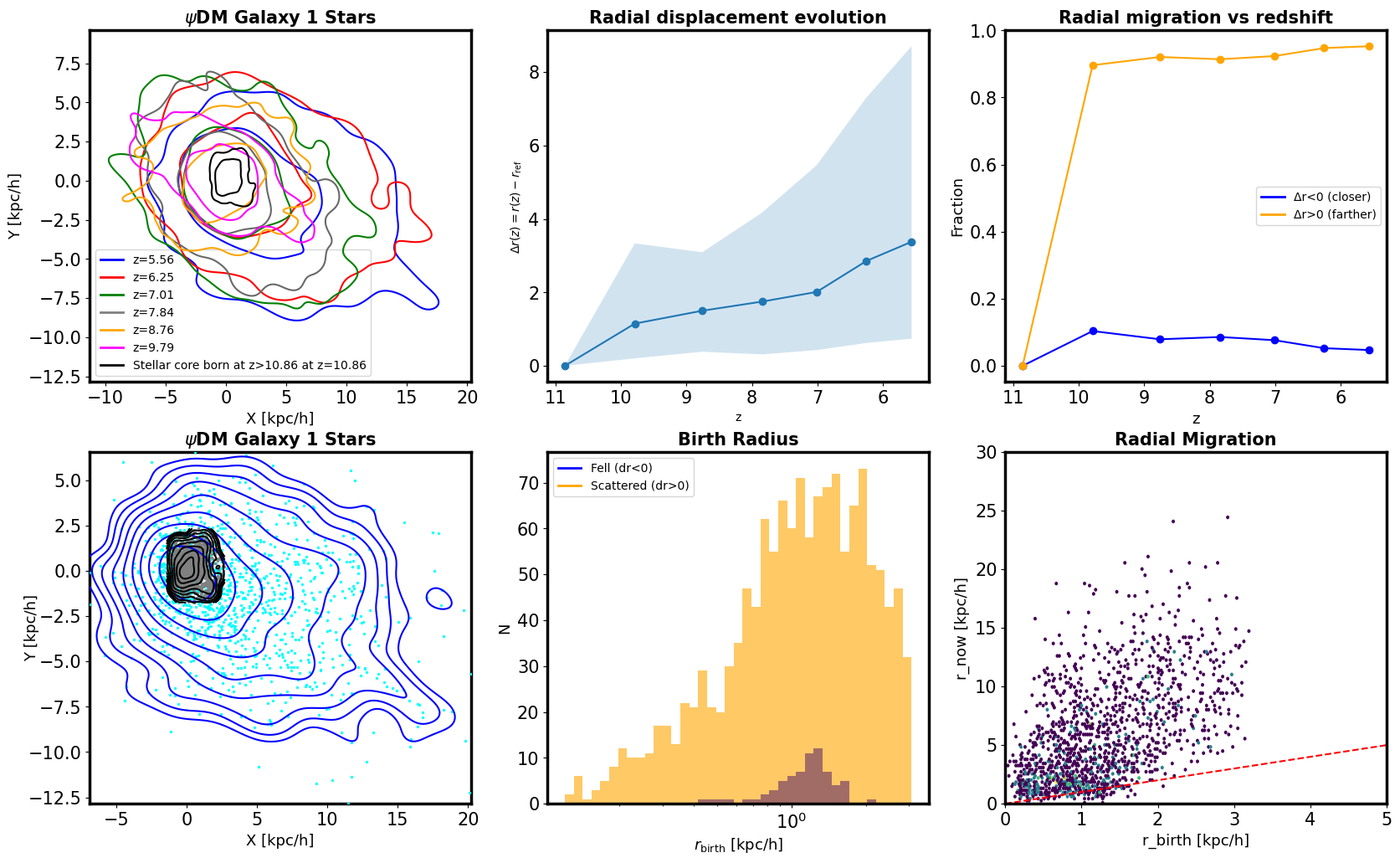}

   \caption{Evolution of the oldest stellar population in the $\psi$DM galaxy, considering only stars already formed at $z\geq10.86$, shown in comoving units. The top-left panel shows the evolution of the stellar density contours from $z=10.86$ to $z=5.56$, illustrating the progressive expansion and diffusion of the initial compact stellar core. The top-middle panel presents the evolution of the mean radial displacement, while the top-right panel shows the fraction of stars migrating inward and outward as a function of redshift, demonstrating that outward diffusion dominates the evolution. The bottom-left panel displays the final stellar density distribution at $z=5.56$, revealing the formation of an extended diffuse stellar halo. The bottom-middle panel compares the birth-radius distributions of inward and outward migrating stars, showing that stars scattered outward predominantly originate from the innermost regions. Finally, the bottom-right panel compares present-day and birth radii, confirming the stochastic outward migration induced by the random walk motion of the central soliton.}
    \label{Fig:CoreScatter}
\end{figure*}

Figure~\ref{Fig:RadialMigration} illustrates the radial migration and stellar scattering processes occurring within the simulated galaxy ( first row is $\psi$DM, second ``WDM'' and third CDM). The first panel compares the birth radius of each star particle, $r_{\rm birth}$, with its present-day radius, $r_{\rm now}$. The red dashed line marks the relation $r_{\rm now}=r_{\rm birth}$, separating stars that migrated outward ($\Delta r > 0$) from those that moved inward ($\Delta r < 0$). For $\psi$DM ,a substantial fraction of stars born in the central regions are found today at significantly larger radii, revealing the existence of strong outward diffusion consistent with a stochastic random-walk process.

The second panel shows the distribution of radial displacements,
\[
\Delta r = r_{\rm now} - r_{\rm birth},
\]
which is centered near zero but exhibits broad asymmetric tails extending toward positive values. This indicates that although some stars migrate inward, the dominant net effect is the gradual scattering of stars toward larger radii. The broad width of the distribution further suggests that the process is highly stochastic rather than driven by smooth secular evolution.

The third and fourth panels compare the birth-radius distributions of stars that migrated inward (``Fell'') and those scattered outward (``Scattered''). The cumulative distribution functions demonstrate that outward-scattered stars preferentially originate from smaller initial radii, whereas stars falling inward tend to be born farther from the center. This behaviour strongly supports a physical picture in which stars formed within the dense central core are progressively extracted and redistributed into the outer halo.

Finally, the boxplot shown in the fifth panel summarizes the statistical properties of the birth radii for both populations, confirming that stars undergoing outward diffusion systematically originate from more central regions. Altogether, these results provide direct evidence for a diffusion-driven redistribution of stellar orbits, naturally explained by the stochastic motion of the central soliton in $\psi$DM halos. Repeated gravitational perturbations induced by the soliton random walk \citep{Schive:2020,Li:2021} gradually increase the orbital energy of stars, populating the outer halo and producing the extended stellar structures observed in diffuse dwarf galaxies and ultra-diffuse systems.

In contrast, neither the ``WDM'' nor the CDM simulations ( second and third rows of figure \ref{Fig:RadialMigration}) exhibit evidence for significant outward stellar scattering. Instead, both models display the opposite behaviour: stars progressively become more centrally concentrated over time as the gravitational potential relaxes and stellar orbits settle toward the halo center. This trend is particularly evident in the radial migration panels, where the stellar distributions remain tightly clustered around the $r_{\rm now}=r_{\rm birth}$ relation, with only limited diffusion toward larger radii. In CDM, the main source of temporary displacement is associated with merger activity at high redshift, but once the halo enters a more quiescent evolutionary phase, the stellar component rapidly reconverges toward the center. Similarly, the ``WDM'' simulation, lacking both mergers and a central soliton \citep{Pozo:2023,Mocz:2020}, shows almost no persistent offset or radial diffusion. These results demonstrate that the strong outward redistribution of stars is unique to $\psi$DM and arises naturally from the stochastic gravitational perturbations generated by the soliton random walk.

\begin{figure*}
   \centering
 \includegraphics[width=1\textwidth,height=9cm]{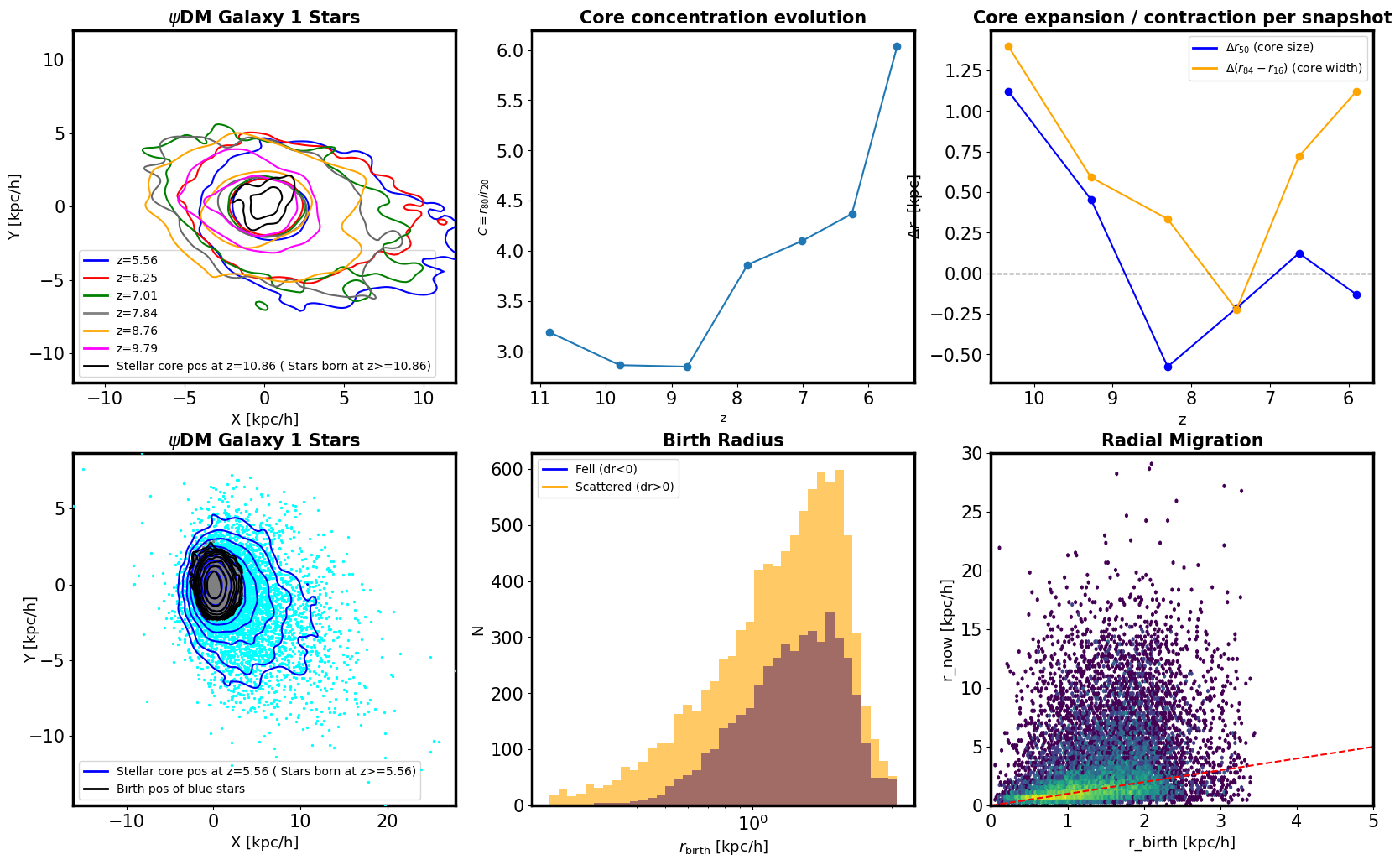}

   \caption{Evolution of the stellar population born within the central core of the $\psi$DM galaxy at all redshifts between $z=10.86$ and $z=5.56$, shown in comoving units. The top-left panel shows the evolution of the stellar density contours, illustrating the gradual expansion of the stellar distribution over time despite the continuous replenishment of the central core by newly formed stars. The top-middle and top-right panels show the evolution of the core concentration and the relative expansion/contraction of the stellar distribution, demonstrating that outward diffusion remains the dominant process. The bottom-left panel displays the final stellar density map at $z=5.56$, revealing the coexistence of a dense central core and an extended diffuse halo. The bottom-middle panel compares the birth-radius distributions of inward and outward migrating stars, while the bottom-right panel shows present-day versus birth radii, confirming that stars formed in the innermost regions are progressively scattered toward larger radii by the stochastic motion of the central soliton.}
    \label{Fig:CoreScatterAll}
\end{figure*}

\begin{figure*}
    \centering
 \includegraphics[width=1\textwidth,height=5cm]{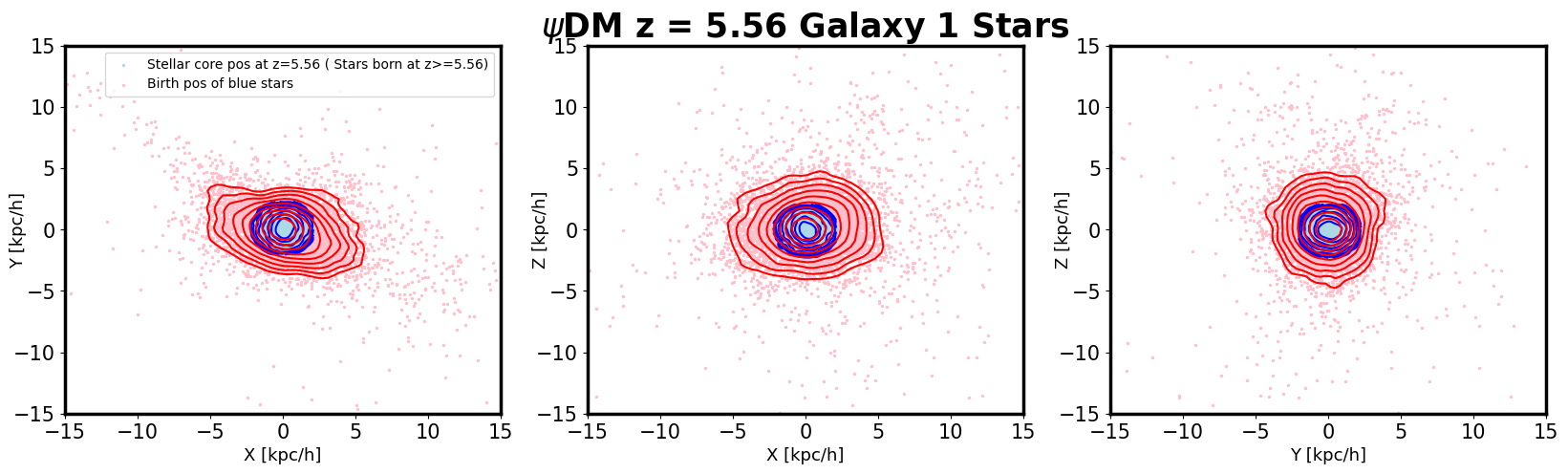}
 \includegraphics[width=1\textwidth,height=4cm]{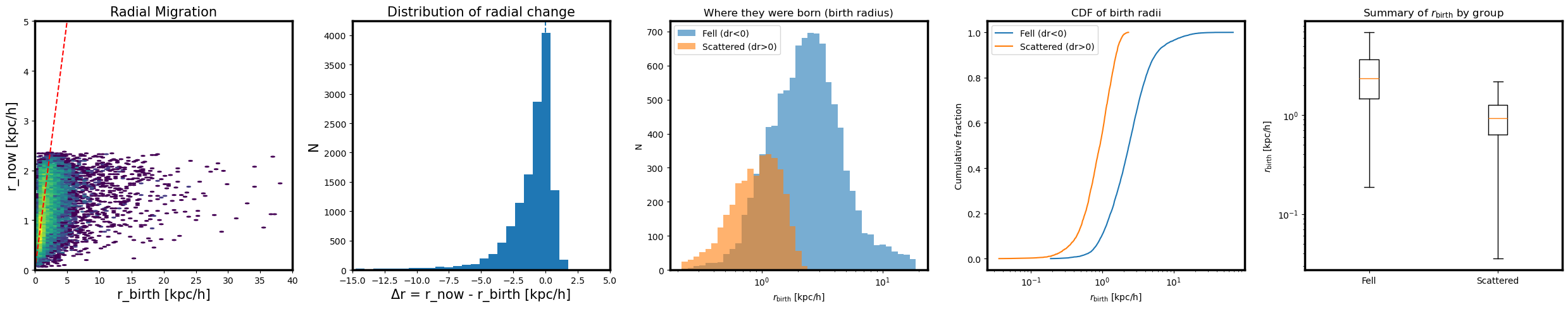}

  \caption{Properties of the stellar population constituting the central core of the $\psi$DM galaxy at $z=5.56$, shown in comoving units. The top panels display the present-day stellar core in three different projections, illustrating the compact central structure surrounded by a diffuse stellar component. The bottom panels characterize the migration properties of these stars through the comparison of present-day and birth radii, the distribution of radial changes, the birth-radius distributions, cumulative distribution functions, and statistical summaries. Most stars currently located in the central core were originally born at larger radii and later migrated inward, while stars born in the innermost regions are preferentially scattered outward. These results support a cyclic picture in which the stellar core is continuously replenished by infalling stars while simultaneously losing stars through stochastic soliton-driven diffusion.}
   \label{Fig:CoreNow}
\end{figure*}

Figure \ref{Fig:CoreScatter} shows the evolution of the stellar core considering only stars that were already formed at $z \geq 10.86$. By excluding stars formed at later times, the figure isolates the dynamical evolution of the oldest stellar population and allows us to directly trace how these early stars are progressively scattered throughout the halo over cosmic time. The first panel presents the projected spatial distribution of the old stellar population at different redshifts. The coloured contours show the evolution of the stellar density from $z=10.86$ down to $z=5.56$. Initially, the stars are tightly concentrated within the central core of the halo. However, as time evolves, the stellar distribution expands significantly, developing a progressively broader and more diffuse morphology. The black contours indicate the stellar core distribution at the initial redshift, highlighting the large increase in spatial extent experienced by the old stellar population. The second panel quantifies the evolution of the mean radial displacement of the stars relative to their initial positions. The average displacement continuously increases with decreasing redshift, demonstrating that stellar orbits undergo cumulative outward diffusion over time. The shaded region marks the dispersion of the radial displacement distribution, showing that the scattering process becomes increasingly stochastic at later times. The third panel shows the fraction of stars that migrate outward ($\Delta r > 0$) or inward ($\Delta r < 0$) as a function of redshift. Shortly after formation, the overwhelming majority of stars become outward migrating particles, while only a small fraction move toward smaller radii. This demonstrates that the net effect of the dynamical evolution is a persistent transfer of stars from the inner core toward the outer halo. The fourth panel displays the final stellar density map at $z=5.56$. The cyan points correspond to the positions of the old stars, while the blue contours trace the projected stellar density. The stellar distribution exhibits a highly diffuse and extended structure surrounding the central core, indicating that stars originally formed in the compact inner regions have been redistributed over several kiloparsecs. The fifth panel compares the birth-radius distributions of stars that migrated inward and outward. Stars that are eventually scattered outward predominantly originate from the innermost regions of the halo, whereas the inward-moving population tends to be born at slightly larger radii. This further supports the interpretation that the central soliton acts as the main driver of the scattering process, extracting stars from the dense core and redistributing them throughout the halo. Finally, the sixth panel compares the present-day radius of each star with its birth radius. The red dashed line marks the relation $r_{\rm now}=r_{\rm birth}$. Most stars lie above this relation, confirming that the dominant evolution is outward migration. The large scatter around the one-to-one relation demonstrates the stochastic nature of the process, consistent with repeated gravitational perturbations induced by the random walk motion of the central soliton in $\psi$DM halos.

Figure \ref{Fig:CoreScatterAll} extends the previous analysis by considering all stars born within the stellar core between $z=10.86$ and $z=5.56$, rather than restricting the sample only to the oldest stellar population formed at $z\geq10.86$. This allows us to evaluate whether the diffusion and scattering processes identified in Figure~\ref{Fig:CoreScatter} remain present when continuous star formation throughout the evolution of the galaxy is taken into account. The overall behaviour remains remarkably consistent with the previous figure. The stellar component continues to evolve from a compact central distribution into a progressively more extended and diffuse structure over time, confirming that the outward migration observed previously is not limited to the oldest stellar population alone. Instead, stars formed at later snapshots are also affected by the same stochastic scattering mechanism generated by the soliton random walk \citep{Schive:2020,Li:2021}. Compared to Figure~\ref{Fig:CoreScatter}, however, the diffusion appears slightly less dramatic due to the continuous injection of newly formed stars into the central regions of the halo. While old stars have had sufficient time to experience repeated gravitational perturbations and migrate to large radii, younger stars remain more centrally concentrated because they have undergone fewer scattering events. As a result, the stellar core is continuously replenished, partially compensating for the outward diffusion of the older populations. This effect is clearly visible in the evolution of the stellar density contours, which still show a substantial expansion of the stellar distribution with time, although with a denser central component than in the previous figure. Similarly, the concentration evolution and radial migration diagnostics indicate that outward scattering remains the dominant dynamical process, with most stars ending at larger radii than their birth positions. The birth-radius distributions also confirm that stars scattered to large distances predominantly originate from the innermost regions of the halo, reinforcing the interpretation that the central soliton acts as the driver of the diffusion process. Altogether, these results demonstrate that soliton-driven stellar diffusion is not an effect restricted to a primordial stellar population, but rather a continuous dynamical mechanism capable of redistributing stars throughout the halo over the entire lifetime of the galaxy. The observed stellar core--halo structure therefore naturally emerges from the competition between ongoing star formation in the central regions and the persistent outward scattering induced by the stochastic motion of the $\psi$DM soliton.

Figure \ref{Fig:CoreNow} further reinforces the interpretation derived from the previous figures by analysing the stellar population that constitutes the central core at $z=5.56$ and tracing back the birth locations of those stars. The upper panels show the present-day stellar core in three different projections, while the lower panels compare the current stellar positions with their birth radii and characterize the migration properties of the population. A key result is that most of the stars currently located within the central core were not originally born there. Instead, the majority formed at significantly larger radii before later migrating inward under the action of gravity. This is clearly visible in the radial migration diagram and in the birth-radius distributions, where the stars classified as ``fell'' ($\Delta r<0$) systematically originate from larger initial radii than those classified as ``scattered'' ($\Delta r>0$). These results strongly support a cyclic dynamical picture for stellar evolution within $\psi$DM halos \citep{Pozo:2020,Pozo:2023,Pozo:2024,Conn:2018,Hui:2021,Broadhurst:2020}. Similar to the behaviour observed in CDM and ``WDM'', stars continuously fall toward the central regions due to gravitational attraction. However, unlike in those models, once stars reach the vicinity of the central soliton they become subject to stochastic gravitational perturbations generated by the soliton random walk \citep{Schive:2020,Li:2021}. These perturbations then scatter a substantial fraction of the stars back toward larger radii, producing a continuous exchange between the inner core and the outer halo. In this framework, the stellar core does not represent a static long-lived structure composed of permanently bound stars, but rather a dynamically evolving region constantly replenished by infalling stars while simultaneously losing stars through soliton-driven diffusion. The extended stellar halos observed in $\psi$DM galaxies \citep{Chiti:2022,Collins:2019,Collins:2021,Pozo:2024,Sato:2025} therefore emerge naturally from the balance between continuous inward gravitational migration and outward scattering induced by the fluctuating soliton potential.

\section{Core-Halo Structure Formation in $\psi$DM}

In this section we discuss how the stellar scattering processes described in the previous sections naturally shape the evolution of the radial stellar profiles in $\psi$DM halos. Our results show that stars born within the central regions, or stars that have remained there for long periods of time, are progressively scattered toward larger radii through repeated gravitational perturbations induced by the soliton random walk. However, despite this continuous outward diffusion, the stellar core is never depleted or destroyed. At the same time that stars are scattered outward (Figures \ref{Fig:CoreScatter}, \ref{Fig:CoreScatterAll}), new stars are continuously formed or migrate inward toward the central regions due to gravitational attraction ( Figure \ref{Fig:CoreNow}). The main consequence of this dynamical equilibrium is the emergence of a flattened stellar core surrounded by an extended diffuse stellar halo, a core--halo structure that is not naturally predicted in standard CDM scenarios but has been widely observed in Local Group dwarf galaxies \citep{Pozo:2020,Pozo:2023,Pozo:20242}. In this framework, the observed stellar structure arises from the balance between the continuous infall of stars toward the center and the outward scattering generated by the fluctuating soliton potential.

\begin{figure*}
   \centering
 \includegraphics[width=1\textwidth,height=9cm]{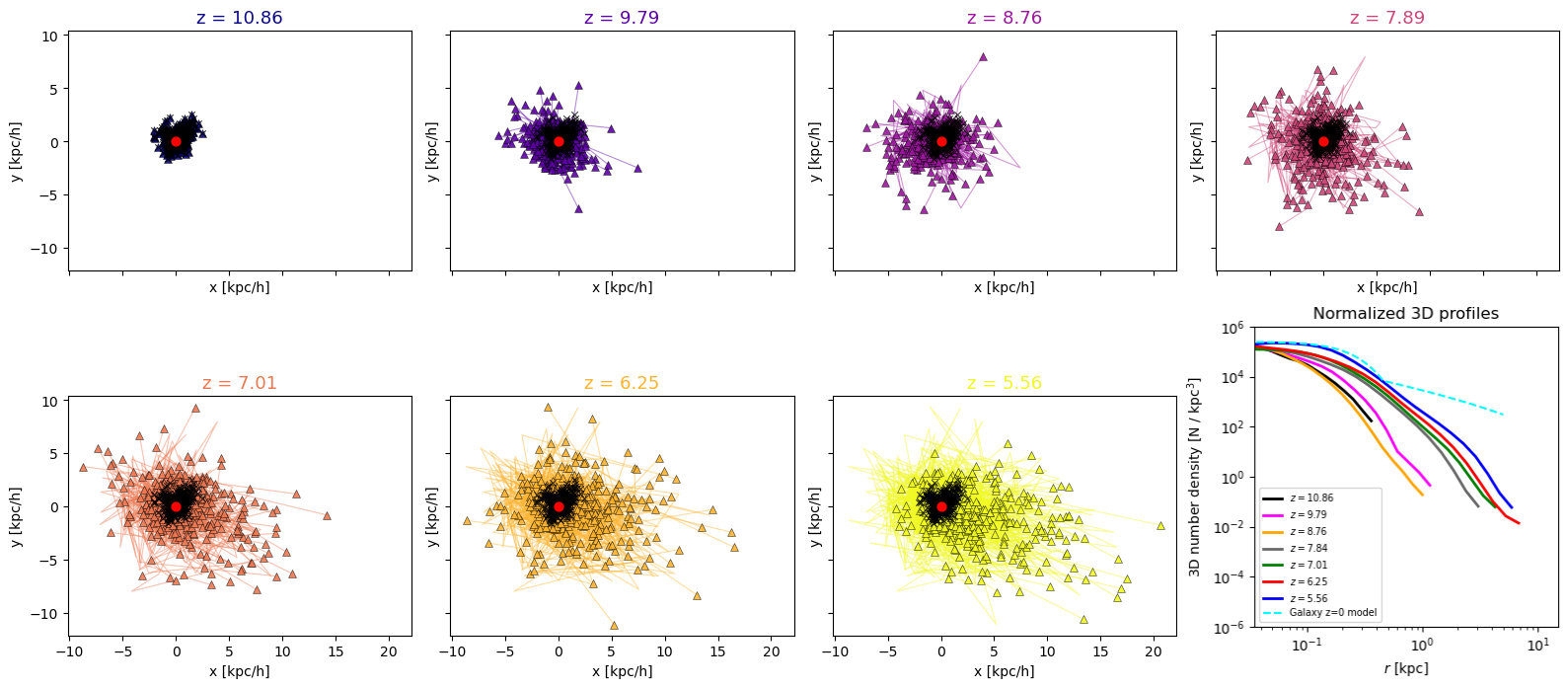}

   \caption{Orbital evolution of stars born in the central core of the $\psi$DM galaxy at $z\geq10.86$, shown in comoving units. Each panel displays the stellar distribution at different redshifts, illustrating the progressive outward diffusion of stellar orbits over time. The black points mark the initial compact stellar core, while the coloured stellar distributions show how stars become increasingly extended and diffuse due to soliton-driven scattering. The final panel presents the evolution of the normalized stellar density profiles, revealing the gradual emergence of a flattened core--halo structure that approaches the diffuse stellar profiles observed in dwarf galaxies. The dashed cyan line shows an idealized core--halo profile representative of observed diffuse systems at $z=0$.}
    \label{Fig:orbits}
\end{figure*}

This behaviour can be clearly seen in the last panel of the second row of Figure~\ref{Fig:orbits}, where we show the evolution of the stellar density profile at every available snapshot. The profile progressively develops a more pronounced core--halo structure with decreasing redshift, converging by $z=5.56$ toward the type of extended stellar profile observationally identified in \cite{Pozo:2023}. For comparison, we include with a dashed cyan line an idealized stellar core--halo profile representative of this class of galaxies at $z=0$ \citep{Pozo:2024,Schive:2016,Schive:20142}. Notice how the inner stellar profile becomes increasingly flatter and broader with time, approaching the characteristic flat core of the idealized profile, while the outer stellar halo continuously increases in density due to the accumulation of scattered stars. The evolution from the earliest to the latest snapshot also reveals a progressive flattening of the outer density slope, indicating the continuous buildup of an extended stellar halo through outward stellar diffusion. The remaining panels of Figure~\ref{Fig:orbits} further illustrate the orbital evolution of stars born at $z\geq10.86$. These panels directly show how stellar orbits gradually expand with time and how the stellar distribution becomes increasingly diffuse relative to the initial compact core marked by the black reference points. Together, these results provide a direct dynamical explanation for the formation of the extended stellar core--halo structures commonly observed in diffuse dwarf galaxies within the Wave Dark Matter framework.

\begin{figure*}
    \centering
 \includegraphics[width=1\textwidth,height=12cm]{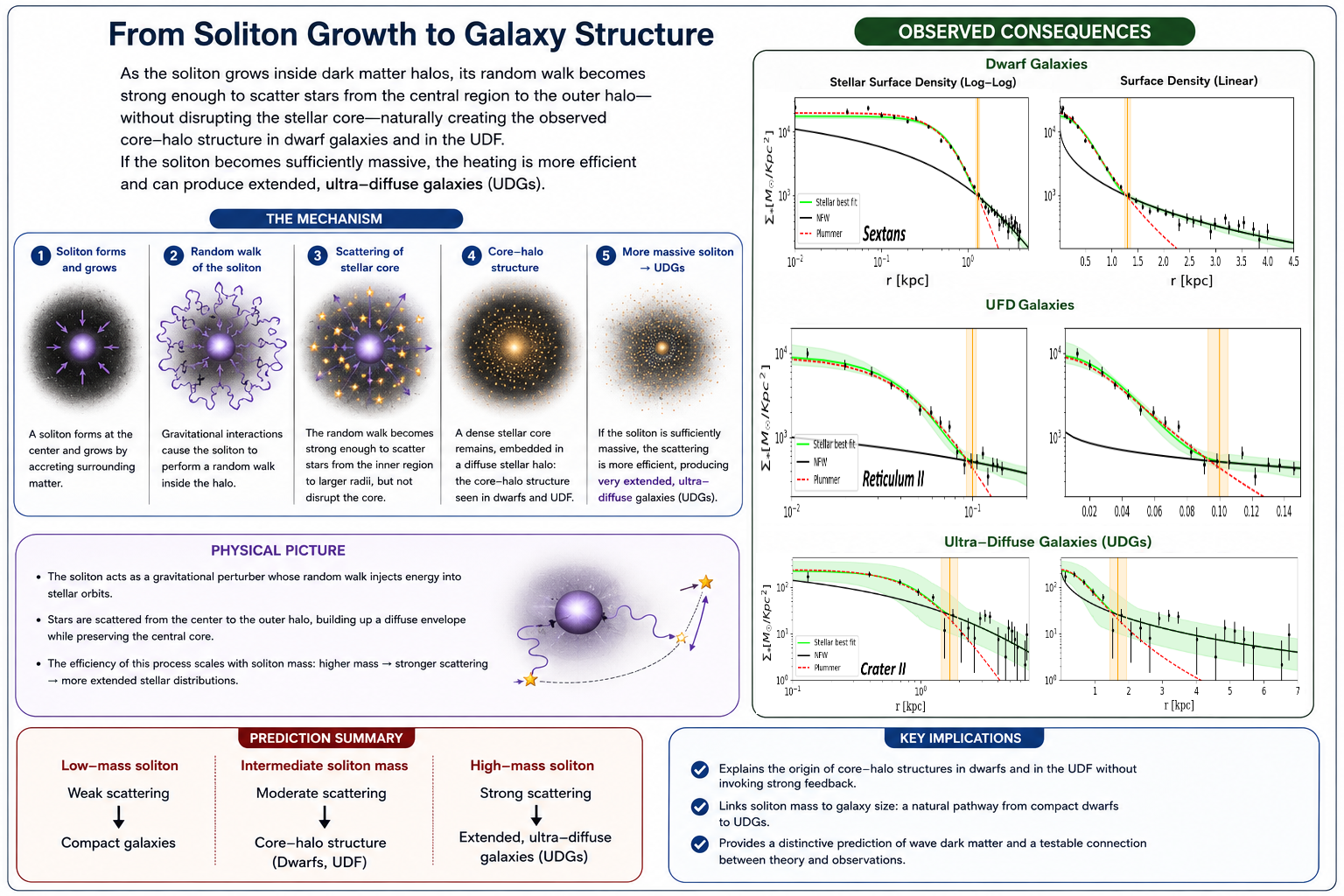}

  \includegraphics[width=1\textwidth,height=11cm]{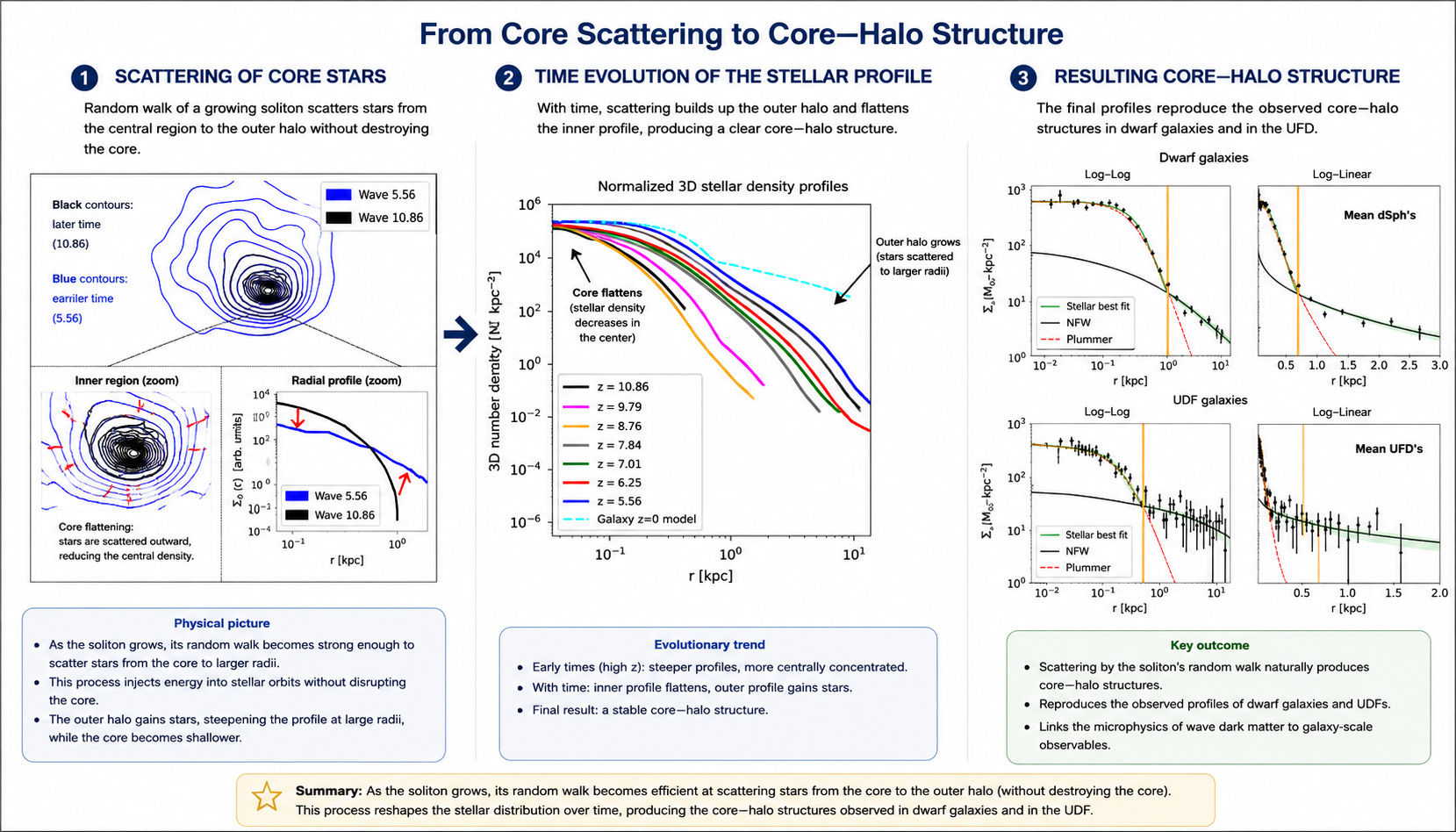}

   \caption{Summary of the soliton-driven stellar diffusion scenario in $\psi$DM halos. The upper panel illustrates how the stochastic random walk of the central soliton scatters stars from the inner core toward the outer halo, naturally producing increasingly diffuse stellar systems from dwarf galaxies to UDGs. The lower panel summarizes the numerical evidence from the simulations, showing the outward migration of stars, the evolution of the stellar density profile toward a core--halo structure, and the agreement with observed dwarf galaxy and UFD stellar profiles.}
    \label{Fig:thesis}
\end{figure*}

We have finally included Figure~\ref{Fig:thesis}, which summarizes the main physical picture and central idea of this work. The figure illustrates how the scattering of the stellar core, a process naturally generated only within Wave Dark Matter halos through the stochastic motion of the central soliton, can explain the universally observed stellar core--halo structures of dwarf galaxies. Depending on the halo mass and the strength of the associated scattering process, this mechanism may account for the stellar distributions observed across a remarkably broad range of systems, from ultra-faint dwarfs (UFDs) hosted by halos of $\sim10^7\,M_\odot$ \citep{Lin:2016,Drlica:2015,Chiti:2021,Martin:2015,Kirby:2015}, through classical dwarf spheroidals, and potentially up to the formation of the extremely diffuse and extended stellar halos observed in ultra-diffuse galaxies (UDGs) with halo masses approaching $\sim10^{11}\,M_\odot$ \citep{Pozo:2021,Gannon:2022,Zaritsky:2023}. Particularly important is the second section of Figure~\ref{Fig:thesis}, labelled ``Scattering of Core Stars'', where we summarize the direct numerical evidence obtained from our simulations. The contour maps show how an initially compact stellar distribution progressively expands and becomes diffuse after prolonged evolution within a $\psi$DM halo. At the same time, the accompanying stellar density profiles demonstrate how stars accumulate within the outer halo as a consequence of this diffusion process. This redistribution is highlighted in the small subpanel by the red arrows, which emphasize the transfer of stellar mass from the central core toward the outer regions of the galaxy.

The possibility that the same mechanism may naturally produce UDG-like systems is briefly introduced in the following section. However, we stress that extending these results toward the UDG regime will require dedicated future simulations covering larger halo masses and longer evolutionary timescales in order to fully test the robustness of this scenario.

\section{Stellar Core Scattering as a Mechanism for UDG Formation}

Ultra-diffuse galaxies (UDGs) constitute a population of low-surface-brightness galaxies characterized by large effective radii and low stellar masses, whose discovery over the past decade has significantly expanded the known diversity of galaxy structural properties \citep{Zaritsky:2022,Zaritsky:2023}. Initially identified in deep imaging surveys of nearby galaxy clusters, UDGs are now known to inhabit a wide range of environments, from rich clusters to galaxy groups and the field. Their extreme combination of size, luminosity, and surface brightness places them at the outskirts of the parameter space occupied by dwarf galaxies, raising fundamental questions about their formation pathways and the nature of their dark-matter haloes\citep{Gannon:2022,Gannon:2024}.

Although UDGs overlap partially with classical dwarf spheroidal galaxies in stellar mass and luminosity, they are readily distinguished by their systematically larger sizes, lower surface brightnesses, and, in some cases, unusual globular cluster populations and internal dynamics \citep{Villaume:2022,Janssens:2022}. These properties challenge simple extrapolations of standard galaxy-formation models and have motivated a wide range of theoretical interpretations, including formation through high-spin dark-matter haloes, tidal heating and stripping, early feedback-driven expansion, and alternative dark-matter scenarios \citep{Rong:2020,Amorisco:2016,DiCintio:2017,Chan:2018,vanDokkum:2018,Robles:2019,Liao:2019,Tremmel:2020}.

Wide-field surveys have now identified tens of thousands of UDG candidates, primarily through automated surface-brightness–limited selections. However, only of order a few thousand systems have been individually studied and characterized in the literature, with detailed measurements of their structural parameters, stellar populations, environments, and, in a smaller subset, internal kinematics. As a result, much of our current understanding of UDGs is built upon a relatively limited sample of well-characterized objects, leaving open key questions regarding their typical dark-matter content, formation histories, and evolutionary connection to the broader dwarf-galaxy population.

In this context, the standard Cold Dark Matter (CDM) paradigm has faced persistent difficulties in accounting for the observed properties and diversity of ultra-diffuse galaxies \citep{vanDokkum:2018,Robles:2019}. UDGs have frequently been interpreted as the outcome of tidal disruption or environmentally driven processes that inhibit efficient galaxy formation, including tidal heating and ram-pressure stripping in dense environments \citep{Liao:2019,Tremmel:2020}. Alternative explanations within the CDM framework have invoked the formation of UDGs in high-spin dark-matter haloes \citep{Amorisco:2016,Rong:2020} or the expansion of stellar distributions through early feedback-driven outflows \citep{DiCintio:2017,Chan:2018}. While these mechanisms can reproduce subsets of the observed population, the discovery of a substantial number of UDGs in low-density and isolated environments poses a significant challenge to scenarios relying primarily on environmental processes. Moreover, several isolated UDGs exhibit modest ongoing star formation and extended neutral hydrogen reservoirs, further complicating a unified explanation within standard CDM models.

In this section, we analyze the UDG population within the $\psi$DM framework, focusing on how the properties of these systems compare with those expected from standard CDM predictions. This approach is motivated by the ability of $\psi$DM to successfully reproduce the internal structure and dynamical properties of dwarf spheroidal galaxies, the closest low-mass analogs to UDGs. 

Figure \ref{Fig:orbits} illustrates how stellar scattering affects the stellar orbits and their final spatial distribution. At the same time, the last panel of the figure, which shows the evolution of the full stellar density profile, reveals the structural consequences of this process. The profile becomes less centrally concentrated due to the continuous scattering of stars from the inner regions. However, the ongoing infall of stars still preserves a central stellar core, consistent with what is observed in dwarf galaxies and UDGs, while simultaneously increasing the size of that core. This effect becomes particularly significant in the presence of a massive soliton, such as those expected in galaxies with halo masses of $\sim 10^{11},M_\odot$, typical of many UDGs. The main consequence of the scattering process is the redistribution of stars toward larger radii, extending the outer stellar profile of the halo. In these extreme cases, the mechanism could naturally generate the very extended and diffuse stellar distributions characteristic of UDGs. For sufficiently massive solitons, where the scattering becomes especially efficient, an initially concentrated galaxy could evolve into a highly diffuse and extended stellar system. This scenario is schematically illustrated in Figure \ref{Fig:thesis}.

\begin{figure*}
    \centering
 \includegraphics[width=1\textwidth,height=6cm]{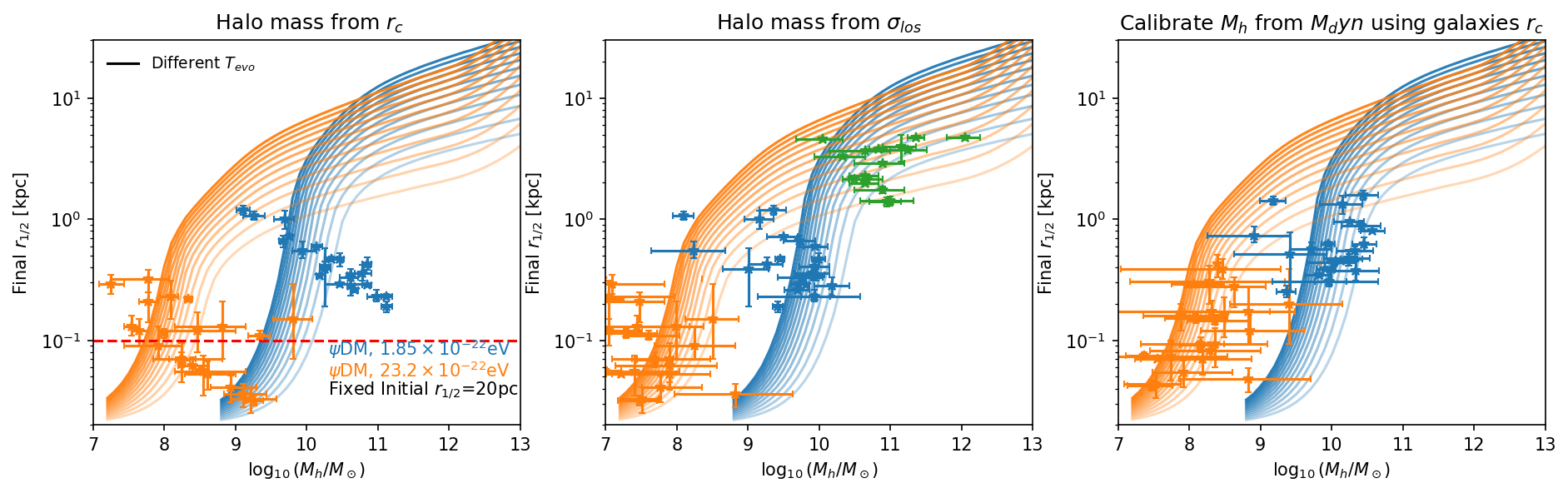}

   \caption{Predicted stellar expansion produced by soliton-driven scattering as a function of halo mass in the $\psi$DM framework. Left panel: halo masses are estimated directly from the core radii reported in \cite{Pozo:20242}, although these measurements become increasingly uncertain for the most compact systems. Middle panel: halo masses are inferred from the observed velocity dispersions using the boson masses preferred from \cite{Pozo:20242} for each galaxy population. Right panel: halo masses are obtained from an empirical calibration between the dynamical mass within the half-light radius and the halo mass, using galaxies with reliable core-radius measurements from \cite{Pozo:20242}.}
    \label{Fig:UDG}
\end{figure*}

Using the scattering framework developed in Zhang et al. 2026, Figure~\ref{Fig:UDG} illustrates the expected stellar expansion induced by soliton-driven scattering as a function of halo mass. The details of the scattering mechanism and the procedure used to compute the evolutionary tracks are described in that work. In all cases, we assume a fixed initial half-light radius of $r_{1/2}=20\,\mathrm{pc}$ and follow its evolution under the action of repeated gravitational encounters with soliton fluctuations.

The blue and orange families of curves represent the idealized evolutionary tracks expected for systems characterized by the boson masses inferred in \cite{Pozo:20242}, corresponding to $m_\psi=1.85\times10^{-22}\,\mathrm{eV}$ and $m_\psi=23.2\times10^{-22}\,\mathrm{eV}$, respectively. Different shades indicate different evolutionary times, spanning $T_{\rm evo}=3$--$14\,\mathrm{Gyr}$. In both cases, more massive halos host more massive solitons and therefore experience more efficient scattering, producing increasingly extended stellar distributions.

The observational data points are computed using the boson mass that provides the best agreement with the location of each galaxy population according to \cite{Pozo:20242}. Their positions are estimated using three different methods to infer the halo mass, shown in the three panels.

In the left panel, the halo mass is obtained directly from the soliton core radius using

\begin{equation}
r_c \propto m_\psi^{-1}M_h^{-1/3},
\end{equation}

where the values of $r_c$ are taken from \citet{Pozo2025}. This approach is subject to significant uncertainties, especially for ultra-faint galaxies. Because of their extremely low luminosities and compact sizes, the half-light radii of UFDs are often only marginally resolved, making their radial profiles considerably less reliable than those of classical dwarf spheroidals. Furthermore, the core radii reported in \cite{Pozo:20242} are themselves approximate quantities inferred from fits to the stellar surface-density profiles rather than exact measurements. The horizontal red dashed line indicates $r_{1/2}=0.1\,\mathrm{kpc}$; below this scale, the inferred core radii become increasingly uncertain and should be interpreted with caution.

In the central panel, the halo mass is estimated from the measured line-of-sight velocity dispersion using

\begin{equation}
M_h \propto \sigma_{\rm los}^{3}\,m_\psi^{-1},
\end{equation}

again adopting the boson masses inferred in \cite{Pozo:20242}. This method avoids the direct use of the poorly constrained core radii and therefore provides a more robust estimate for compact systems.

Finally, in the right panel we calibrate an empirical relation between the dynamical mass enclosed within the half-light radius and the halo mass. The dynamical mass is computed using the Wolf et al.~\citep{Wolf2010} estimator,

\begin{equation}
M_{\rm dyn}=930\,\sigma_{\rm los}^{2}R_e,
\end{equation}

and we fit a power-law relation of the form

\begin{equation}
M_h=A\,M_{\rm dyn}^{\alpha},
\end{equation}

using galaxies with reliable core-radius measurements. This calibration enables halo masses to be inferred even for systems lacking direct $r_c$ estimates, providing an independent consistency check of the soliton-scattering scenario.

Although the tracks shown in Figure~\ref{Fig:UDG} should be regarded as idealized evolutionary sequences, the overall agreement between the predicted expansion and the observed distribution of ultra-faint dwarfs and dwarf spheroidals suggests that soliton-driven scattering may play an important role in shaping the structural diversity of diffuse galaxies.

\section{Conclusions}

In this work we have explored the dynamical consequences of soliton random-walk motion in Wave Dark Matter ($\psi$DM) halos and its impact on the evolution of stellar distributions in low-mass galaxies. Using cosmological simulations comparing $\psi$DM, CDM, and ``WDM'' scenarios, we find that the stochastic motion of the central soliton naturally induces repeated gravitational perturbations that progressively scatter stars from the dense inner regions toward the outer halo. This process produces a continuous outward diffusion of stellar orbits that is absent in both CDM and ``WDM'' simulations.

Our analysis demonstrates that stars born in the central regions of $\psi$DM halos experience a gradual increase in orbital radius over cosmic time, generating extended and diffuse stellar structures. At the same time, the stellar core itself is continuously replenished by newly formed stars and by stars migrating inward under gravity. The resulting balance between inward migration and outward scattering naturally produces a long-lived stellar core--halo structure, characterized by a flattened central stellar density profile surrounded by an extended diffuse stellar halo.

The simulations further show that this mechanism depends strongly on halo mass through the corresponding growth of the soliton. In more massive halos, the soliton becomes more massive and dynamically influential, increasing the efficiency of stellar scattering and producing progressively more diffuse stellar systems. Under a fixed boson mass of $m_\psi = 1.8 \times 10^{-22},\mathrm{eV}$, the predicted stellar expansion follows the observed mass--size relation spanning ultra-faint dwarfs (UFDs), dwarf spheroidals (dSphs), and ultra-diffuse galaxies (UDGs). This suggests that these apparently distinct galaxy populations may instead represent different manifestations of the same underlying wave-driven diffusion process.

Importantly, the $\psi$DM simulations reproduce behaviours that are not observed in CDM or ``WDM''. In the latter models, stellar populations become increasingly concentrated with time, except during temporary merger-driven disturbances. By contrast, the persistent stochastic displacement of the soliton in $\psi$DM continuously perturbs stellar orbits even in isolated systems, naturally generating diffuse stellar envelopes without requiring strong feedback processes, tidal stripping, or unusually high halo spin parameters.

Our results therefore provide a unified physical framework capable of explaining the widespread presence of stellar core--halo structures in dwarf galaxies and potentially the origin of ultra-diffuse galaxies. In this picture, UDGs are not necessarily failed massive galaxies or systems requiring exceptional environmental histories, but may emerge naturally from enhanced stellar diffusion within massive $\psi$DM halos. The observed diversity of galaxy sizes and surface brightness profiles could then arise primarily from the interplay between halo mass, soliton strength, and evolutionary time.

Future work will require larger cosmological simulations extending to more massive halos and lower redshifts in order to fully test the UDG regime and compare directly with the rapidly growing observational samples from surveys such as \textit{Euclid}. Nevertheless, the results presented here strongly suggest that soliton-driven stellar diffusion may constitute one of the key observable signatures of the intrinsic granular dynamics of Wave Dark Matter.

\begin{acknowledgements} 
We warmly acknowledge Douglas Finkbeiner for very fruitful conversations. A.P. is grateful for the continued support of the DIPC postgraduate program as well as the Center for Astrophysics | Harvard \& Smithsonian for their warm hospitality at the beginning of this project. R.E. acknowledges the support from grant numbers 21-atp21-0077, NSF AST-1816420, and HST-GO-16173.001-A as well as the Institute for Theory and Computation at the Center for Astrophysics. We are grateful to the supercomputer facility at Harvard University where most of the simulation work was done. G.S. is grateful to the IAS at HKUST for their generous support. TJB supported by the Spanish grant PID2023-149016NBI00 (funded by MCIN/AEI/10.13039/501100011033 andby “ERDF A way of making Europe”. P.M. acknowledges this work was in part performed under the auspices of the U.S. Department of Energy by Lawrence Livermore National Laboratory under contract DE-AC52-07NA27344, Lawrence Livermore National Security, LLC.

\end{acknowledgements} 

\section*{Data availability}

The data underlying this article will be shared on reasonable
request to the corresponding author.

\bibliographystyle{aa} 
\bibliography{apssamp}

@PREAMBLE{
 "\providecommand{\noopsort}[1]{}" 
 # "\providecommand{\singleletter}[1]{#1}%" 
}

@ARTICLE{ Kirby:2015,
   author = {Kirby, E, N. and Simon, J. D. and Cohen, J. G.},
   title = { Spectroscopic Confirmation of the Dwarf Galaxies Hydra II and Pisces II and the Globular Cluster Laevens 1.},
   journal = {Astrophys. J.},
   volume = {810},
   pages = {12},
   year = {2015},
   doi={ https://doi.org/10.1088/0004-637X/810/1/56 }
}

@ARTICLE{Pozo:2023,
       author = {{Pozo}, A. and {Emami}, R. and {Mocz}, P. and {Broadhurst}, T. and {Hernquist}, L. and {Vogelsberger}, M. and {Smith}, R. and {Tremblay}, G. and {Narayan}, R. and {Steiner}, J. and {Grindlay}, J. and {Smoot}, G.},
        title = "{Galaxy formation with wave/fuzzy dark matter: The core-halo structure and the solitonic imprint}",
      journal = {\aap},
         year = 2025,
        month = jul,
       volume = {699},
          eid = {A308},
        pages = {A308},
          doi = {10.1051/0004-6361/202450443},
archivePrefix = {arXiv},
       eprint = {2310.12217},
 primaryClass = {astro-ph.CO},
       adsurl = {https://ui.adsabs.harvard.edu/abs/2025A&A...699A.308P}
}

@ARTICLE{Mocz:2020,
   author = { Mocz, P. and others},
   title = {Galaxy formation with BECDM - II. Cosmic filaments and first galaxies.},
   journal = {Mon. Not. R. Astron. Soc.},
   volume = {494},
   pages = {2027-2044},
   year = {2020},
   doi={ https://doi.org/10.1093/mnras/staa738   }
}

@ARTICLE{Mocz:2019,
       author = {{Mocz}, Philip and {Fialkov}, Anastasia and {Vogelsberger}, Mark and {Becerra}, Fernando and {Amin}, Mustafa A. and {Bose}, Sownak and {Boylan-Kolchin}, Michael and {Chavanis}, Pierre-Henri and {Hernquist}, Lars and {Lancaster}, Lachlan and {Marinacci}, Federico and {Robles}, Victor H. and {Zavala}, Jes{\'u}s},
        title = "{First Star-Forming Structures in Fuzzy Cosmic Filaments}",
      journal = {\prl},
         year = 2019,
        month = oct,
       volume = {123},
       number = {14},
          eid = {141301},
        pages = {141301},
          doi = {10.1103/PhysRevLett.123.141301},
archivePrefix = {arXiv},
       eprint = {1910.01653},
 primaryClass = {astro-ph.GA},
       adsurl = {https://ui.adsabs.harvard.edu/abs/2019PhRvL.123n1301M}
}

@ARTICLE{Collins:2019,
   author = {  Collins, M. L. M. and Tollerud, E. J. and Rich, R. and Ibata, R. A. and Martin, N. F. and Chapman, S. C. and Gilbert, K. M. and Preston, J. },
   title = {A detailed study of Andromeda XIX, an extreme local analogue of ultra diffuse galaxies.},
   journal = {Mon. Not. R. Astron. Soc.},
   volume = {491},
   pages = {3496-3514},
   year = {2019},
   doi={ https://doi.org/10.1093/mnras/stz3252 }
}

@ARTICLE{Planck:2016,
   author = {Planck Collaboration},
   title = {Planck 2015 results. XIII. Cosmological parameters.},
   journal = {Astro. $\&$ Astrophys. },
   volume = {594},
   pages = {63},
   year = {2016},
   doi={ https://doi.org/10.1051/0004-6361/201525830 }
}

@ARTICLE{ Mocz:2017,
   author = {Mocz, P. and Vogelsberger, M. and Robles, V. H. and Zavala, J. and Boylan-Kolchin, M. and Fialkov, A. and Hernquist, L. },
   title = {Galaxy formation with BECDM - I. Turbulence and relaxation of idealized haloes.},
   journal = {Mon. Not. R. Astron. Soc.},
   volume = {471},
   pages = {4559-4570},
   year = {2017},
   doi={ https://doi.org/10.1093/mnras/stx1887   }
}

@ARTICLE{Broadhurst:2020,
   author = { Broadhurst, T. and  de Martino, I. and  Luu, H. N. and  Smoot, G. F. and  Tye, S. -H. H.},
   title = {Ghostly galaxies as solitons of Bose-Einstein dark matter.},
   journal = {Phys. Rev. D.},
   volume = {101},
    year = {2020},
   doi={ https://doi.org/10.1103/PhysRevD.101.083012 }
}

@ARTICLE{Collins:2021,
   author = { Collins, M. L. M. and Read, J. I. and Ibata, R. A. and Rich, R. M. and Martin, N. F. and Peñarrubia, J. and Chapman, S. C. and Tollerud, E. J. and Weisz, D. R. },
   title = { Andromeda XXI -- a dwarf galaxy in a low density dark matter halo
},
   journal = {ArXiv},
   year = {2021},
   doi={https://arxiv.org/pdf/2102.11890.pdf}
}

@ARTICLE{Hu:2000,
   author = { Hu, W. and Barkana, R. and  Gruzinov, A.},
   title = {Fuzzy Cold Dark Matter: The Wave Properties of Ultralight Particles.},
   journal = {Phys. Rev. Lett.},
   volume = {85},
   pages = {1158-1161},
   year = {2000},
   doi={ https://doi.org/10.1103/PhysRevLett.85.1158}
}

@ARTICLE{Safarzadeh:2021,
   author = {  Safarzadeh, M. and Loeb, A.},
   title = {A New Challenge for Dark Matter Models.},
   journal = {Arxiv},
 
   year = {2021},
   doi={https://arxiv.org/pdf/2107.03478.pdf }
}

@ARTICLE{ Klypin:1999,
   author = {Klypin, A. and Kravtsov, A. V. and Valenzuela, O. Prada, F.},
   title = {Where Are the Missing Galactic Satellites?},
   journal = {Astrophys. J.},
   volume = {522},
   pages = {82-92},
   year = {1999},
   doi={ https://doi.org/10.1086/307643 }
}

@ARTICLE{ Moore:1994,
   author = { Moore, B.},
   title = {Evidence against dissipation-less dark matter from observations of galaxy haloes.},
   journal = {Nature},
   volume = {370},
   pages = {629-631},
   year = {1994},
   doi={ https://doi.org/10.1038/370629a0 }
}

@ARTICLE{Navarro:1996,
       author = {{Navarro}, Julio F. and {Frenk}, Carlos S. and {White}, Simon D.~M.},
        title = "{The Structure of Cold Dark Matter Halos}",
      journal = {\apj},
         year = 1996,
        month = may,
       volume = {462},
        pages = {563},
          doi = {10.1086/177173},
archivePrefix = {arXiv},
       eprint = {astro-ph/9508025},
 primaryClass = {astro-ph},
       adsurl = {https://ui.adsabs.harvard.edu/abs/1996ApJ...462..563N}
}

@ARTICLE{ Pozo:2020,
   author = {  Pozo, A. and Broadhurst, T. and de Martino, I. and Luu, H. N. and Smoot, George F. and Lim, J. and Neyrinck, M.},
   title = {Wave Dark Matter and Ultra Diffuse Galaxies.},
   journal = {Mon. Not. R. Astron. Soc.},
   volume = {504},
   pages = {2868-2876},
   year = {2020},
   doi={ https://doi.org/10.1093/mnras/stab855   }
}

@ARTICLE{Schive:2016,
       author = {{Schive}, Hsi-Yu and {Chiueh}, Tzihong and {Broadhurst}, Tom and {Huang}, Kuan-Wei},
        title = "{Contrasting Galaxy Formation from Quantum Wave Dark Matter, {\ensuremath{\psi}}DM, with {\ensuremath{\Lambda}}CDM, using Planck and Hubble Data}",
      journal = {\apj},
         year = 2016,
        month = feb,
       volume = {818},
       number = {1},
          eid = {89},
        pages = {89},
          doi = {10.3847/0004-637X/818/1/89},
archivePrefix = {arXiv},
       eprint = {1508.04621},
 primaryClass = {astro-ph.GA},
       adsurl = {https://ui.adsabs.harvard.edu/abs/2016ApJ...818...89S}
}

@ARTICLE{ Widrow:1993,
   author = {Widrow, L. M. and Kaiser, N.},
   title = { Using the Schroedinger Equation to Simulate Collisionless Matter.},
   journal = {Astrophys. J.},
   volume = {416},
   pages = {L71-L74},
   year = {1993},
   doi={ https://doi.org/10.1086/187073 }
}

@ARTICLE{Woo:2009,
       author = {{Woo}, Tak-Pong and {Chiueh}, Tzihong},
        title = "{High-Resolution Simulation on Structure Formation with Extremely Light Bosonic Dark Matter}",
      journal = {\apj},
         year = 2009,
        month = may,
       volume = {697},
       number = {1},
        pages = {850-861},
          doi = {10.1088/0004-637X/697/1/850},
archivePrefix = {arXiv},
       eprint = {0806.0232},
 primaryClass = {astro-ph},
       adsurl = {https://ui.adsabs.harvard.edu/abs/2009ApJ...697..850W}
}

@ARTICLE{ May:2021,
   author = {May, S. and Springel, V.},
   title = { Structure formation in large-volume cosmological simulations of fuzzy dark matter: Impact of the non-linear dynamics.},
   journal = {Mon. Not. R. Astron. Soc.},
   volume = {Advanced access},
   year = {2021},
   doi={ https://doi.org/10.1093/mnras/stab1764 }
 }

@ARTICLE{Chiti:2021,
       author = {{Chiti}, Anirudh and {Frebel}, Anna and {Simon}, Joshua D. and {Erkal}, Denis and {Chang}, Laura J. and {Necib}, Lina and {Ji}, Alexander P. and {Jerjen}, Helmut and {Kim}, Dongwon and {Norris}, John E.},
        title = "{An extended halo around an ancient dwarf galaxy}",
      journal = {Nature Astronomy},
         year = 2021,
        month = apr,
       volume = {5},
        pages = {392-400},
          doi = {10.1038/s41550-020-01285-w},
archivePrefix = {arXiv},
       eprint = {2012.02309},
 primaryClass = {astro-ph.GA},
       adsurl = {https://ui.adsabs.harvard.edu/abs/2021NatAs...5..392C}
}

@ARTICLE{Chiti:2022,
       author = {{Chiti}, Anirudh and {Frebel}, Anna and {Ji}, Alexander P. and {Mardini}, Mohammad K. and {Ou}, Xiaowei and {Simon}, Joshua D. and {Jerjen}, Helmut and {Kim}, Dongwon and {Norris}, John E.},
        title = "{Detailed Chemical Abundances of Stars in the Outskirts of the Tucana II Ultrafaint Dwarf Galaxy}",
      journal = {\aj},
         year = 2023,
        month = feb,
       volume = {165},
       number = {2},
          eid = {55},
        pages = {55},
          doi = {10.3847/1538-3881/aca416},
archivePrefix = {arXiv},
       eprint = {2205.01740},
 primaryClass = {astro-ph.GA},
       adsurl = {https://ui.adsabs.harvard.edu/abs/2023AJ....165...55C}
}

@ARTICLE{Pozo:2022,
       author = {{Pozo}, Alvaro and {Broadhurst}, Tom and {Emami}, Razieh and {Smoot}, George},
        title = "{Understanding the 'feeble giant' Crater II with tidally stretched wave dark matter}",
      journal = {\mnras},
         year = 2022,
        month = sep,
       volume = {515},
       number = {2},
        pages = {2624-2632},
          doi = {10.1093/mnras/stac1862},
archivePrefix = {arXiv},
       eprint = {2112.06514},
 primaryClass = {astro-ph.CO},
       adsurl = {https://ui.adsabs.harvard.edu/abs/2022MNRAS.515.2624P}
}

@ARTICLE{Schive:2014,
       author = {{Schive}, Hsi-Yu and {Chiueh}, Tzihong and {Broadhurst}, Tom},
        title = "{Cosmic structure as the quantum interference of a coherent dark wave}",
      journal = {Nature Physics},
         year = 2014,
        month = jul,
       volume = {10},
       number = {7},
        pages = {496-499},
          doi = {10.1038/nphys2996},
archivePrefix = {arXiv},
       eprint = {1406.6586},
 primaryClass = {astro-ph.GA},
       adsurl = {https://ui.adsabs.harvard.edu/abs/2014NatPh..10..496S}
}

@ARTICLE{Schive:20142,
       author = {{Schive}, Hsi-Yu and {Liao}, Ming-Hsuan and {Woo}, Tak-Pong and {Wong}, Shing-Kwong and {Chiueh}, Tzihong and {Broadhurst}, Tom and {Hwang}, W. -Y. Pauchy},
        title = "{Understanding the Core-Halo Relation of Quantum Wave Dark Matter from 3D Simulations}",
      journal = {\prl},
         year = 2014,
        month = dec,
       volume = {113},
       number = {26},
          eid = {261302},
        pages = {261302},
          doi = {10.1103/PhysRevLett.113.261302},
archivePrefix = {arXiv},
       eprint = {1407.7762},
 primaryClass = {astro-ph.GA},
       adsurl = {https://ui.adsabs.harvard.edu/abs/2014PhRvL.113z1302S}
}

@ARTICLE{ Schwabe:2016,
   author = { Schwabe, B. and Niemeyer, J. C. and Engels, J. F.},
   title = { Simulations of solitonic core mergers in ultralight axion dark matter cosmologies.},
   journal = {Phys. Rev. D.},
   volume = {94},
  
   year = {2016},
   doi={ https://doi.org/ 10.1103/PhysRevD.94.043513    }
}

@ARTICLE{ Veltmaat:2018,
   author = {Veltmaat, J. and Niemeyer, J. C. and Schwabe, B.},
   title = { Formation and structure of ultralight bosonic dark matter halos.},
   journal = {Phys. Rev. D.},
   volume = {98},
    
   year = {2018},
   doi={ https://doi.org/10.1103/PhysRevD.98.043509 }
   }

@ARTICLE{Hui:2017,
       author = {{Hui}, Lam and {Ostriker}, Jeremiah P. and {Tremaine}, Scott and {Witten}, Edward},
        title = "{Ultralight scalars as cosmological dark matter}",
      journal = {\prd},
         year = 2017,
        month = feb,
       volume = {95},
       number = {4},
          eid = {043541},
        pages = {043541},
          doi = {10.1103/PhysRevD.95.043541},
archivePrefix = {arXiv},
       eprint = {1610.08297},
 primaryClass = {astro-ph.CO},
       adsurl = {https://ui.adsabs.harvard.edu/abs/2017PhRvD..95d3541H}
}

@ARTICLE{Lin:2016,
   author = { Lin, W. and Ishak, M.},
   title = {Ultra faint dwarf galaxies: an arena for testing dark matter versus modified gravity.},
   journal = {Journal of Cosmo. and Astroparticle Phy.},
    year = {2016},
   doi={ https://doi.org/10.1088/1475-7516/2016/10/025}
}

@ARTICLE{Conn:2018,
       author = {{Conn}, Blair C. and {Jerjen}, Helmut and {Kim}, Dongwon and {Schirmer}, Mischa},
        title = "{On the Nature of Ultra-faint Dwarf Galaxy Candidates. I. DES1, Eridanus III, and Tucana V}",
      journal = {\apj},
         year = 2018,
        month = jan,
       volume = {852},
       number = {2},
          eid = {68},
        pages = {68},
          doi = {10.3847/1538-4357/aa9eda},
archivePrefix = {arXiv},
       eprint = {1712.01439},
 primaryClass = {astro-ph.GA},
       adsurl = {https://ui.adsabs.harvard.edu/abs/2018ApJ...852...68C}
}

@ARTICLE{Martin:2015,
       author = {{Martin}, Nicolas F. and {Nidever}, David L. and {Besla}, Gurtina and {Olsen}, Knut and {Walker}, Alistair R. and {Vivas}, A. Katherina and {Gruendl}, Robert A. and {Kaleida}, Catherine C. and {Mu{\~n}oz}, Ricardo R. and {Blum}, Robert D. and {Saha}, Abhijit and {Conn}, Blair C. and {Bell}, Eric F. and {Chu}, You-Hua and {Cioni}, Maria-Rosa L. and {de Boer}, Thomas J.~L. and {Gallart}, Carme and {Jin}, Shoko and {Kunder}, Andrea and {Majewski}, Steven R. and {Martinez-Delgado}, David and {Monachesi}, Antonela and {Monelli}, Matteo and {Monteagudo}, Lara and {No{\"e}l}, Noelia E.~D. and {Olszewski}, Edward W. and {Stringfellow}, Guy S. and {van der Marel}, Roeland P. and {Zaritsky}, Dennis},
        title = "{Hydra II: A Faint and Compact Milky Way Dwarf Galaxy Found in the Survey of the Magellanic Stellar History}",
      journal = {\apjl},
         year = 2015,
        month = may,
       volume = {804},
       number = {1},
          eid = {L5},
        pages = {L5},
          doi = {10.1088/2041-8205/804/1/L5},
archivePrefix = {arXiv},
       eprint = {1503.06216},
 primaryClass = {astro-ph.GA},
       adsurl = {https://ui.adsabs.harvard.edu/abs/2015ApJ...804L...5M}
}

@ARTICLE{Padmanabhan:2021,
   author = {Padmanabhan, H. and Loeb, A. },
   title = {Distinguishing AGN from starbursts as the origin of double-peaked Lyman-alpha emitters in the reionization era.},
   journal = {Astron. Astrophys.},
   volume = {646},
   pages = {4},
   year = {2021},
   doi={ https://doi.org/10.1051/0004-6361/202040107 }
}

@ARTICLE{Drlica:2015,
       author = {{Drlica-Wagner}, A. and {Bechtol}, K. and {Rykoff}, E.~S. and {Luque}, E. and {Queiroz}, A. and {Mao}, Y. -Y. and {Wechsler}, R.~H. and {Simon}, J.~D. and {Santiago}, B. and {Yanny}, B. and {Balbinot}, E. and {Dodelson}, S. and {Fausti Neto}, A. and {James}, D.~J. and {Li}, T.~S. and {Maia}, M.~A.~G. and {Marshall}, J.~L. and {Pieres}, A. and {Stringer}, K. and {Walker}, A.~R. and {Abbott}, T.~M.~C. and {Abdalla}, F.~B. and {Allam}, S. and {Benoit-L{\'e}vy}, A. and {Bernstein}, G.~M. and {Bertin}, E. and {Brooks}, D. and {Buckley-Geer}, E. and {Burke}, D.~L. and {Carnero Rosell}, A. and {Carrasco Kind}, M. and {Carretero}, J. and {Crocce}, M. and {da Costa}, L.~N. and {Desai}, S. and {Diehl}, H.~T. and {Dietrich}, J.~P. and {Doel}, P. and {Eifler}, T.~F. and {Evrard}, A.~E. and {Finley}, D.~A. and {Flaugher}, B. and {Fosalba}, P. and {Frieman}, J. and {Gaztanaga}, E. and {Gerdes}, D.~W. and {Gruen}, D. and {Gruendl}, R.~A. and {Gutierrez}, G. and {Honscheid}, K. and {Kuehn}, K. and {Kuropatkin}, N. and {Lahav}, O. and {Martini}, P. and {Miquel}, R. and {Nord}, B. and {Ogando}, R. and {Plazas}, A.~A. and {Reil}, K. and {Roodman}, A. and {Sako}, M. and {Sanchez}, E. and {Scarpine}, V. and {Schubnell}, M. and {Sevilla-Noarbe}, I. and {Smith}, R.~C. and {Soares-Santos}, M. and {Sobreira}, F. and {Suchyta}, E. and {Swanson}, M.~E.~C. and {Tarle}, G. and {Tucker}, D. and {Vikram}, V. and {Wester}, W. and {Zhang}, Y. and {Zuntz}, J. and {DES Collaboration}},
        title = "{Eight Ultra-faint Galaxy Candidates Discovered in Year Two of the Dark Energy Survey}",
      journal = {\apj},
         year = 2015,
        month = nov,
       volume = {813},
       number = {2},
          eid = {109},
        pages = {109},
          doi = {10.1088/0004-637X/813/2/109},
archivePrefix = {arXiv},
       eprint = {1508.03622},
 primaryClass = {astro-ph.GA},
       adsurl = {https://ui.adsabs.harvard.edu/abs/2015ApJ...813..109D}
}

@ARTICLE{Marsh:2014,
   author = {Marsh, D. J. E. and Silk, J.},
   title = {A model for halo formation with axion mixed dark matter.},
   journal = {Mon. Not. R. Astron. Soc.},
   volume = {437},
   pages = {2652-2663},
   year = {2014},
   doi={ https://doi.org/10.1093/mnras/stt2079    }
}

@ARTICLE{Zaritsky:2023,
       author = {{Zaritsky}, Dennis and {Donnerstein}, Richard and {Dey}, Arjun and {Karunakaran}, Ananthan and {Kadowaki}, Jennifer and {Khim}, Donghyeon J. and {Spekkens}, Kristine and {Zhang}, Huanian},
        title = "{Systematically Measuring Ultra-diffuse Galaxies (SMUDGes). V. The Complete SMUDGes Catalog and the Nature of Ultradiffuse Galaxies}",
      journal = {\apjs},
         year = 2023,
        month = aug,
       volume = {267},
       number = {2},
          eid = {27},
        pages = {27},
          doi = {10.3847/1538-4365/acdd71},
archivePrefix = {arXiv},
       eprint = {2306.01524},
 primaryClass = {astro-ph.GA},
       adsurl = {https://ui.adsabs.harvard.edu/abs/2023ApJS..267...27Z}
}

@ARTICLE{Zaritsky:2022,
       author = {{Zaritsky}, Dennis and {Donnerstein}, Richard and {Karunakaran}, Ananthan and {Barbosa}, C.~E. and {Dey}, Arjun and {Kadowaki}, Jennifer and {Spekkens}, Kristine and {Zhang}, Huanian},
        title = "{Systematically Measuring Ultra-diffuse Galaxies (SMUDGes). III. The Southern SMUDGes Catalog}",
      journal = {\apjs},
         year = 2022,
        month = aug,
       volume = {261},
       number = {2},
          eid = {11},
        pages = {11},
          doi = {10.3847/1538-4365/ac6ceb},
archivePrefix = {arXiv},
       eprint = {2205.02193},
 primaryClass = {astro-ph.GA},
       adsurl = {https://ui.adsabs.harvard.edu/abs/2022ApJS..261...11Z}
}

@ARTICLE{Gannon:2022,
       author = {{Gannon}, Jonah S. and {Forbes}, Duncan A. and {Romanowsky}, Aaron J. and {Ferr{\'e}-Mateu}, Anna and {Couch}, Warrick J. and {Brodie}, Jean P. and {Huang}, Song and {Janssens}, Steven R. and {Okabe}, Nobuhiro},
        title = "{Ultra-diffuse galaxies in the perseus cluster: comparing galaxy properties with globular cluster system richness}",
      journal = {\mnras},
         year = 2022,
        month = feb,
       volume = {510},
       number = {1},
        pages = {946-958},
          doi = {10.1093/mnras/stab3297},
archivePrefix = {arXiv},
       eprint = {2111.06007},
 primaryClass = {astro-ph.GA},
       adsurl = {https://ui.adsabs.harvard.edu/abs/2022MNRAS.510..946G}
}

@ARTICLE{Gannon:2024,
       author = {{Gannon}, Jonah S. and {Ferr{\'e}-Mateu}, Anna and {Forbes}, Duncan A. and {Brodie}, Jean P. and {Buzzo}, Maria Luisa and {Romanowsky}, Aaron J.},
        title = "{A Catalogue and analysis of ultra-diffuse galaxy spectroscopic properties}",
      journal = {\mnras},
         year = 2024,
        month = jun,
       volume = {531},
       number = {1},
        pages = {1856-1869},
          doi = {10.1093/mnras/stae1287},
archivePrefix = {arXiv},
       eprint = {2405.09104},
 primaryClass = {astro-ph.GA},
       adsurl = {https://ui.adsabs.harvard.edu/abs/2024MNRAS.531.1856G}
}

@ARTICLE{Villaume:2022,
       author = {{Villaume}, Alexa and {Romanowsky}, Aaron J. and {Brodie}, Jean and {van Dokkum}, Pieter and {Conroy}, Charlie and {Forbes}, Duncan A. and {Danieli}, Shany and {Martin}, Christopher and {Matuszewski}, Matt},
        title = "{Spatially Resolved Stellar Spectroscopy of the Ultra-diffuse Galaxy Dragonfly 44. III. Evidence for an Unexpected Star Formation History under Conventional Galaxy Evolution Processes}",
      journal = {\apj},
         year = 2022,
        month = jan,
       volume = {924},
       number = {1},
          eid = {32},
        pages = {32},
          doi = {10.3847/1538-4357/ac341e},
archivePrefix = {arXiv},
       eprint = {2101.02220},
 primaryClass = {astro-ph.GA},
       adsurl = {https://ui.adsabs.harvard.edu/abs/2022ApJ...924...32V}
}

@ARTICLE{Janssens:2022,
       author = {{Janssens}, Steven R. and {Romanowsky}, Aaron J. and {Abraham}, Roberto and {Brodie}, Jean P. and {Couch}, Warrick J. and {Forbes}, Duncan A. and {Laine}, Seppo and {Mart{\'\i}nez-Delgado}, David and {van Dokkum}, Pieter G.},
        title = "{The globular clusters and star formation history of the isolated, quiescent ultra-diffuse galaxy DGSAT I}",
      journal = {\mnras},
         year = 2022,
        month = nov,
       volume = {517},
       number = {1},
        pages = {858-871},
          doi = {10.1093/mnras/stac2717},
archivePrefix = {arXiv},
       eprint = {2209.09910},
 primaryClass = {astro-ph.GA},
       adsurl = {https://ui.adsabs.harvard.edu/abs/2022MNRAS.517..858J}
}

@ARTICLE{Amorisco:2016,
       author = {{Amorisco}, N.~C. and {Loeb}, A.},
        title = "{Ultradiffuse galaxies: the high-spin tail of the abundant dwarf galaxy population}",
      journal = {\mnras},
         year = 2016,
        month = jun,
       volume = {459},
       number = {1},
        pages = {L51-L55},
          doi = {10.1093/mnrasl/slw055},
archivePrefix = {arXiv},
       eprint = {1603.00463},
 primaryClass = {astro-ph.GA},
       adsurl = {https://ui.adsabs.harvard.edu/abs/2016MNRAS.459L..51A}
}

@ARTICLE{Rong:2020,
       author = {{Rong}, Yu and {Mancera Pi{\~n}a}, Pavel E. and {Tempel}, Elmo and {Puzia}, Thomas H. and {De Rijcke}, Sven},
        title = "{Exploring the origin of ultra-diffuse galaxies in clusters from their primordial alignment}",
      journal = {\mnras},
         year = 2020,
        month = nov,
       volume = {498},
       number = {1},
        pages = {L72-L76},
          doi = {10.1093/mnrasl/slaa129},
archivePrefix = {arXiv},
       eprint = {2007.06593},
 primaryClass = {astro-ph.GA},
       adsurl = {https://ui.adsabs.harvard.edu/abs/2020MNRAS.498L..72R}
}

@ARTICLE{vanDokkum:2018,
       author = {{van Dokkum}, Pieter and {Danieli}, Shany and {Cohen}, Yotam and {Merritt}, Allison and {Romanowsky}, Aaron J. and {Abraham}, Roberto and {Brodie}, Jean and {Conroy}, Charlie and {Lokhorst}, Deborah and {Mowla}, Lamiya and {O'Sullivan}, Ewan and {Zhang}, Jielai},
        title = "{A galaxy lacking dark matter}",
      journal = {\nat},
         year = 2018,
        month = mar,
       volume = {555},
       number = {7698},
        pages = {629-632},
          doi = {10.1038/nature25767},
archivePrefix = {arXiv},
       eprint = {1803.10237},
 primaryClass = {astro-ph.GA},
       adsurl = {https://ui.adsabs.harvard.edu/abs/2018Natur.555..629V}
}

@ARTICLE{DiCintio:2017,
       author = {{Di Cintio}, Arianna and {Brook}, Chris B. and {Dutton}, Aaron A. and {Macci{\`o}}, Andrea V. and {Obreja}, Aura and {Dekel}, Avishai},
        title = "{NIHAO - XI. Formation of ultra-diffuse galaxies by outflows}",
      journal = {\mnras},
         year = 2017,
        month = mar,
       volume = {466},
       number = {1},
        pages = {L1-L6},
          doi = {10.1093/mnrasl/slw210},
archivePrefix = {arXiv},
       eprint = {1608.01327},
 primaryClass = {astro-ph.GA},
       adsurl = {https://ui.adsabs.harvard.edu/abs/2017MNRAS.466L...1D}
}

@ARTICLE{Liao:2019,
       author = {{Liao}, Shihong and {Gao}, Liang and {Frenk}, Carlos S. and {Grand}, Robert J.~J. and {Guo}, Qi and {G{\'o}mez}, Facundo A. and {Marinacci}, Federico and {Pakmor}, R{\"u}diger and {Shao}, Shi and {Springel}, Volker},
        title = "{Ultra-diffuse galaxies in the Auriga simulations}",
      journal = {\mnras},
         year = 2019,
        month = dec,
       volume = {490},
       number = {4},
        pages = {5182-5195},
          doi = {10.1093/mnras/stz2969},
archivePrefix = {arXiv},
       eprint = {1904.06356},
 primaryClass = {astro-ph.GA},
       adsurl = {https://ui.adsabs.harvard.edu/abs/2019MNRAS.490.5182L}
}

@INPROCEEDINGS{Tremmel:2020,
       author = {{Tremmel}, M. and {Wright}, A. and {Brooks}, A. and {Munshi}, F. and {Nagai}, D. and {Quinn}, T.},
        title = "{The Formation of Ultra-Diffuse Galaxies from Passive Evolution in the RomulusC Galaxy Cluster Simulation}",
    booktitle = {American Astronomical Society Meeting Abstracts \#235},
         year = 2020,
       series = {American Astronomical Society Meeting Abstracts},
       volume = {235},
        month = jan,
          eid = {316.02},
        pages = {316.02},
       adsurl = {https://ui.adsabs.harvard.edu/abs/2020AAS...23531602T}
}

@ARTICLE{Robles:2019,
       author = {{Robles}, Victor H. and {Bullock}, James S. and {Boylan-Kolchin}, Michael},
        title = "{Scalar field dark matter: helping or hurting small-scale problems in cosmology?}",
      journal = {\mnras},
         year = 2019,
        month = feb,
       volume = {483},
       number = {1},
        pages = {289-298},
          doi = {10.1093/mnras/sty3190},
archivePrefix = {arXiv},
       eprint = {1807.06018},
 primaryClass = {astro-ph.CO},
       adsurl = {https://ui.adsabs.harvard.edu/abs/2019MNRAS.483..289R}
}

@ARTICLE{Chan:2018,
       author = {{Chan}, T.~K. and {Kere{\v{s}}}, D. and {Wetzel}, A. and {Hopkins}, P.~F. and {Faucher-Gigu{\`e}re}, C.-A. and {El-Badry}, K. and {Garrison-Kimmel}, S. and {Boylan-Kolchin}, M.},
        title = "{The origin of ultra diffuse galaxies: stellar feedback and quenching}",
      journal = {\mnras},
         year = 2018,
        month = jul,
       volume = {478},
       number = {1},
        pages = {906-925},
          doi = {10.1093/mnras/sty1153},
archivePrefix = {arXiv},
       eprint = {1711.04788},
 primaryClass = {astro-ph.GA},
       adsurl = {https://ui.adsabs.harvard.edu/abs/2018MNRAS.478..906C}
}

@ARTICLE{Pozo:2021,
       author = {{Pozo}, Alvaro and {Broadhurst}, Tom and {de Martino}, Ivan and {Luu}, Hoang Nhan and {Smoot}, George F. and {Lim}, Jeremy and {Neyrinck}, Mark},
        title = "{Wave dark matter and ultra-diffuse galaxies}",
      journal = {\mnras},
         year = 2021,
        month = jun,
       volume = {504},
       number = {2},
        pages = {2868-2876},
          doi = {10.1093/mnras/stab855},
archivePrefix = {arXiv},
       eprint = {2003.08313},
 primaryClass = {astro-ph.GA},
       adsurl = {https://ui.adsabs.harvard.edu/abs/2021MNRAS.504.2868P}
}

@ARTICLE{Schive:2020,
       author = {{Schive}, Hsi-Yu and {Chiueh}, Tzihong and {Broadhurst}, Tom},
        title = "{Soliton Random Walk and the Cluster-Stripping Problem in Ultralight Dark Matter}",
      journal = {\prl},
         year = 2020,
        month = may,
       volume = {124},
       number = {20},
          eid = {201301},
        pages = {201301},
          doi = {10.1103/PhysRevLett.124.201301},
archivePrefix = {arXiv},
       eprint = {1912.09483},
 primaryClass = {astro-ph.GA},
       adsurl = {https://ui.adsabs.harvard.edu/abs/2020PhRvL.124t1301S}
}

@ARTICLE{Li:2021,
       author = {{Li}, Xinyu and {Hui}, Lam and {Yavetz}, Tomer D.},
        title = "{Oscillations and random walk of the soliton core in a fuzzy dark matter halo}",
      journal = {\prd},
         year = 2021,
        month = jan,
       volume = {103},
       number = {2},
          eid = {023508},
        pages = {023508},
          doi = {10.1103/PhysRevD.103.023508},
archivePrefix = {arXiv},
       eprint = {2011.11416},
 primaryClass = {astro-ph.CO},
       adsurl = {https://ui.adsabs.harvard.edu/abs/2021PhRvD.103b3508L}
}

@ARTICLE{deBlok:2010,
       author = {{de Blok}, W.~J.~G.},
        title = "{The Core-Cusp Problem}",
      journal = {Advances in Astronomy},
         year = 2010,
        month = jan,
       volume = {2010},
          eid = {789293},
        pages = {789293},
          doi = {10.1155/2010/789293},
archivePrefix = {arXiv},
       eprint = {0910.3538},
 primaryClass = {astro-ph.CO},
       adsurl = {https://ui.adsabs.harvard.edu/abs/2010AdAst2010E...5D}
}

@ARTICLE{Pozo:2025,
       author = {{Pozo}, A. and {Emami}, R. and {Mocz}, P. and {Broadhurst}, T. and {Hernquist}, L. and {Vogelsberger}, M. and {Smith}, R. and {Tremblay}, G. and {Narayan}, R. and {Steiner}, J. and {Grindlay}, J. and {Smoot}, G.},
        title = "{Galaxy formation with wave/fuzzy dark matter: The core-halo structure and the solitonic imprint}",
      journal = {\aap},
         year = 2025,
        month = jul,
       volume = {699},
          eid = {A308},
        pages = {A308},
          doi = {10.1051/0004-6361/202450443},
archivePrefix = {arXiv},
       eprint = {2310.12217},
 primaryClass = {astro-ph.CO},
       adsurl = {https://ui.adsabs.harvard.edu/abs/2025A&A...699A.308P}
}

@ARTICLE{Pozo:2024,
       author = {{Pozo}, Alvaro and {Broadhurst}, Tom and {de Martino}, Ivan and {Chiueh}, Tzihong and {Smoot}, George F. and {Bonoli}, Silvia and {Angulo}, Raul},
        title = "{Detection of a universal core-halo transition in dwarf galaxies as predicted by Bose-Einstein dark matter}",
      journal = {\prd},
         year = 2024,
        month = aug,
       volume = {110},
       number = {4},
          eid = {043534},
        pages = {043534},
          doi = {10.1103/PhysRevD.110.043534},
archivePrefix = {arXiv},
       eprint = {2010.10337},
 primaryClass = {astro-ph.GA},
       adsurl = {https://ui.adsabs.harvard.edu/abs/2024PhRvD.110d3534P}
}

@ARTICLE{Pozo:20242,
       author = {{Pozo}, Alvaro and {Broadhurst}, Tom and {Smoot}, George F. and {Chiueh}, Tzihong and {Luu}, Hoang Nhan and {Vogelsberger}, Mark and {Mocz}, Philip},
        title = "{Dwarf galaxies united by dark bosons}",
      journal = {\prd},
         year = 2024,
        month = apr,
       volume = {109},
       number = {8},
          eid = {083532},
        pages = {083532},
          doi = {10.1103/PhysRevD.109.083532},
archivePrefix = {arXiv},
       eprint = {2302.00181},
 primaryClass = {astro-ph.CO},
       adsurl = {https://ui.adsabs.harvard.edu/abs/2024PhRvD.109h3532P}
}

@ARTICLE{Hui:2021,
       author = {{Hui}, Lam},
        title = "{Wave Dark Matter}",
      journal = {\araa},
         year = 2021,
        month = sep,
       volume = {59},
        pages = {247-289},
          doi = {10.1146/annurev-astro-120920-010024},
archivePrefix = {arXiv},
       eprint = {2101.11735},
 primaryClass = {astro-ph.CO},
       adsurl = {https://ui.adsabs.harvard.edu/abs/2021ARA&A..59..247H}
}

@ARTICLE{Lin:2018,
       author = {{Lin}, Shan-Chang and {Schive}, Hsi-Yu and {Wong}, Shing-Kwong and {Chiueh}, Tzihong},
        title = "{Self-consistent construction of virialized wave dark matter halos}",
      journal = {\prd},
         year = 2018,
        month = may,
       volume = {97},
       number = {10},
          eid = {103523},
        pages = {103523},
          doi = {10.1103/PhysRevD.97.103523},
archivePrefix = {arXiv},
       eprint = {1801.02320},
 primaryClass = {astro-ph.CO},
       adsurl = {https://ui.adsabs.harvard.edu/abs/2018PhRvD..97j3523L}
}

@ARTICLE{Madelung:1927,
       author = {{Madelung}, E.},
        title = "{Quantentheorie in hydrodynamischer Form}",
      journal = {Zeitschrift fur Physik},
         year = 1927,
        month = mar,
       volume = {40},
       number = {3-4},
        pages = {322-326},
          doi = {10.1007/BF01400372},
       adsurl = {https://ui.adsabs.harvard.edu/abs/1927ZPhy...40..322M}
}

@ARTICLE{Naoz:2012,
       author = {{Naoz}, Smadar and {Yoshida}, Naoki and {Gnedin}, Nickolay Y.},
        title = "{Simulations of Early Baryonic Structure Formation with Stream Velocity. I. Halo Abundance}",
      journal = {\apj},
         year = 2012,
        month = mar,
       volume = {747},
       number = {2},
          eid = {128},
        pages = {128},
          doi = {10.1088/0004-637X/747/2/128},
archivePrefix = {arXiv},
       eprint = {1108.5176},
 primaryClass = {astro-ph.CO},
       adsurl = {https://ui.adsabs.harvard.edu/abs/2012ApJ...747..128N}
}

@ARTICLE{Pandya:2024,
       author = {{Pandya}, Viraj and {Zhang}, Haowen and {Huertas-Company}, Marc and {Iyer}, Kartheik G. and {McGrath}, Elizabeth and {Barro}, Guillermo and {Finkelstein}, Steven L. and {K{\"u}mmel}, Martin and {Hartley}, William G. and {Ferguson}, Henry C. and {Kartaltepe}, Jeyhan S. and {Primack}, Joel and {Dekel}, Avishai and {Faber}, Sandra M. and {Koo}, David C. and {Bryan}, Greg L. and {Somerville}, Rachel S. and {Amor{\'\i}n}, Ricardo O. and {Arrabal Haro}, Pablo and {Bagley}, Micaela B. and {Bell}, Eric F. and {Bertin}, Emmanuel and {Costantin}, Luca and {Dav{\'e}}, Romeel and {Dickinson}, Mark and {Feldmann}, Robert and {Fontana}, Adriano and {Gavazzi}, Raphael and {Giavalisco}, Mauro and {Grazian}, Andrea and {Grogin}, Norman A. and {Guo}, Yuchen and {Hahn}, ChangHoon and {Holwerda}, Benne W. and {Kewley}, Lisa J. and {Kirkpatrick}, Allison and {Kocevski}, Dale D. and {Koekemoer}, Anton M. and {Lotz}, Jennifer M. and {Lucas}, Ray A. and {Papovich}, Casey and {Pentericci}, Laura and {P{\'e}rez-Gonz{\'a}lez}, Pablo G. and {Pirzkal}, Nor and {Ravindranath}, Swara and {Rose}, Caitlin and {Schefer}, Marc and {Simons}, Raymond C. and {Straughn}, Amber N. and {Tacchella}, Sandro and {Trump}, Jonathan R. and {de la Vega}, Alexander and {Wilkins}, Stephen M. and {Wuyts}, Stijn and {Yang}, Guang and {Yung}, L.~Y. Aaron},
        title = "{Galaxies Going Bananas: Inferring the 3D Geometry of High-redshift Galaxies with JWST-CEERS}",
      journal = {\apj},
         year = 2024,
        month = mar,
       volume = {963},
       number = {1},
          eid = {54},
        pages = {54},
          doi = {10.3847/1538-4357/ad1a13},
archivePrefix = {arXiv},
       eprint = {2310.15232},
 primaryClass = {astro-ph.GA},
       adsurl = {https://ui.adsabs.harvard.edu/abs/2024ApJ...963...54P}
}

@ARTICLE{Sato:2025,
       author = {{Sato}, Kyosuke S. and {Okamoto}, Sakurako and {Yagi}, Masafumi and {Komiyama}, Yutaka and {Arimoto}, Nobuo and {Wyse}, Rosemary F.~G. and {Kirby}, Evan N. and {Chiba}, Masashi and {Ogami}, Itsuki and {Tanaka}, Mikito},
        title = "{The Extended Stellar Distribution in the Outskirts of the Ursa Minor Dwarf Spheroidal Galaxy}",
      journal = {\apjl},
         year = 2025,
        month = nov,
       volume = {993},
       number = {1},
          eid = {L7},
        pages = {L7},
          doi = {10.3847/2041-8213/ae0cb3},
archivePrefix = {arXiv},
       eprint = {2509.20914},
 primaryClass = {astro-ph.GA},
       adsurl = {https://ui.adsabs.harvard.edu/abs/2025ApJ...993L...7S}
}

\end{document}